\documentclass[twocolumn]{aastex631}
\shorttitle{El Gordo as seen by JWST}
\shortauthors{Diego et al.}

\usepackage{amsmath}
\usepackage{txfonts}
\usepackage[T1]{fontenc}
\usepackage{longtable}
\newcommand{\vectX}{\bf {\it X}}
\newcommand{\vectTheta}{\bf {\it \Theta}}
\newcommand{\matrGamma}{\bf \Gamma}

\newcommand{\EG}{El~Gordo}
\newcommand{\Msol}{\hbox{M$_\odot$}}
\newcommand{\Lsol}{\hbox{L$_\odot$}}
\newcommand{\no}{\nodata}

\begin{document}

\title{JWST's PEARLS: a new lens model for ACT-CL J0102$-$4915, ``\EG'', and the first red supergiant star at cosmological distances discovered by JWST}

\correspondingauthor{Jose M. Diego} \email{jdiego@ifca.unican.es}


\author[0000-0001-9065-3926]{Jose M. Diego}
\affiliation{Instituto de F\'isica de Cantabria (CSIC-UC). Avda. Los Castros s/n. 39005 Santander, Spain}  

\author[0000-0002-7876-4321]{Ashish K. Meena}
\affiliation{Physics Department, Ben-Gurion University of the Negev, P.O. Box 653, Beer-Sheva 8410501, Israel}

\author[0000-0003-4875-6272]{Nathan J. Adams} 
\affiliation{Jodrell Bank Centre for Astrophysics, Alan Turing Building, 
University of Manchester, Oxford Road, Manchester M13 9PL, UK}

\author[0000-0002-5807-4411]{Tom Broadhurst} 
\affiliation{Department of Theoretical Physics, University of the Basque
Country UPV-EHU, 48040 Bilbao, Spain}
\affiliation{Donostia International Physics Center (DIPC), 20018 Donostia, The
Basque Country}
\affiliation{IKERBASQUE, Basque Foundation for Science, Alameda Urquijo, 36-5
48008 Bilbao, Spain}

\author[0000-0003-2091-8946]{Liang Dai} 
\affiliation{Department of Physics, 366 Physics North MC 7300, University of California, Berkeley, CA 94720, USA}

\author[0000-0001-7410-7669]{Dan Coe} 
\affiliation{AURA for the European Space Agency (ESA), Space Telescope Science
Institute, 3700 San Martin Drive, Baltimore, MD 21210, USA}

\author[0000-0003-1625-8009]{Brenda Frye} 
\affiliation{Steward Observatory, University of Arizona, 933 N Cherry Ave,  
Tucson, AZ, 85721-0009, USA}

\author[0000-0003-3142-997X]{Patrick Kelly} 
\affiliation{School of Physics and Astronomy, University of Minnesota, 116 Church Street SE, Minneapolis, MN 55455, USA} 

\author[0000-0002-6610-2048]{Anton M. Koekemoer} 
\affiliation{Space Telescope Science Institute, 3700 San Martin Drive, Baltimore, MD 21210, USA}

\author[0000-0002-2282-8795]{Massimo Pascale} 
\affiliation{Department of Astronomy, University of California, 501 Campbell
Hall \#3411, Berkeley, CA 94720, USA}

\author[0000-0002-9895-5758]{S.~P.~Willner} 
\affiliation{Center for Astrophysics \textbar\ Harvard \& Smithsonian, 60 Garden St., Cambridge, MA 02138, USA}

\author[0000-0003-1096-2636]{Erik Zackrisson} 
\affiliation{Observational Astrophysics, Department of Physics and Astronomy, Uppsala University, Box 516, SE-751 20 Uppsala, Sweden}

\author[0000-0002-0350-4488]{Adi Zitrin} 
\affiliation{Physics Department, Ben-Gurion University of the Negev, P.O. Box
653, Beer-Sheva 8410501, Israel}

\author[0000-0001-8156-6281]{Rogier A. Windhorst}
\affiliation{School of Earth and Space Exploration, Arizona State University,
Tempe, AZ 85287-1404, USA}

\author[0000-0003-3329-1337]{Seth H. Cohen} 
\affiliation{School of Earth and Space Exploration, Arizona State University,
Tempe, AZ 85287-1404, USA}

\author[0000-0003-1268-5230]{Rolf A. Jansen} 
\affiliation{School of Earth and Space Exploration, Arizona State University,
Tempe, AZ 85287-1404, USA}

\author[0000-0002-7265-7920]{Jake Summers} 
\affiliation{School of Earth and Space Exploration, Arizona State University,
Tempe, AZ 85287-1404, USA}

\author[0000-0001-9052-9837]{Scott Tompkins} 
\affiliation{School of Earth and Space Exploration, Arizona State University,
Tempe, AZ 85287-1404, USA}

\author[0000-0003-1949-7638]{Christopher J. Conselice} 
\affiliation{Jodrell Bank Centre for Astrophysics, Alan Turing Building, 
University of Manchester, Oxford Road, Manchester M13 9PL, UK}

\author[0000-0001-9491-7327]{Simon P. Driver} 
\affiliation{International Centre for Radio Astronomy Research (ICRAR) and the
International Space Centre (ISC), The University of Western Australia, M468,
35 Stirling Highway, Crawley, WA 6009, Australia}

\author[0000-0001-7592-7714]{Haojing Yan} 
\affiliation{Department of Physics and Astronomy, University of Missouri,
Columbia, MO 65211, USA}



\author[0000-0001-9440-8872]{Norman Grogin} 
\affiliation{Space Telescope Science Institute, 3700 San Martin Drive, Baltimore, MD 21210, USA}


\author[0000-0001-6434-7845]{Madeline A. Marshall} 
\affiliation{National Research Council of Canada, Herzberg Astronomy \&
Astrophysics Research Centre, 5071 West Saanich Road, Victoria, BC V9E 2E7, 
Canada}
\affiliation{ARC Centre of Excellence for All Sky Astrophysics in 3 Dimensions
(ASTRO 3D), Australia}

\affiliation{National Research Council of Canada, Herzberg Astronomy \&
Astrophysics Research Centre, 5071 West Saanich Road, Victoria, BC V9E 2E7,
Canada; \& ARC Centre of Excellence for All Sky Astrophysics in 3
Dimensions (ASTRO 3D), Australia}

\author[0000-0003-3382-5941]{Nor Pirzkal} 
\affiliation{Space Telescope Science Institute, 3700 San Martin Drive,
Baltimore, MD 21210, USA}

\author[0000-0003-0429-3579]{Aaron Robotham} 
\affiliation{International Centre for Radio Astronomy Research (ICRAR) and the
International Space Centre (ISC), The University of Western Australia, M468,
35 Stirling Highway, Crawley, WA 6009, Australia}

\author[0000-0003-0894-1588]{Russell E. Ryan, Jr.} 
\affiliation{Space Telescope Science Institute, 3700 San Martin Drive, 
Baltimore, MD 21210, USA}

\author[0000-0001-9262-9997]{Christopher N. A. Willmer} 
\affiliation{Steward Observatory, University of Arizona, 933 N Cherry Ave,
Tucson, AZ, 85721-0009, USA}

\author[0000-0002-7908-9284]{Larry D. Bradley}
\affiliation{Space Telescope Science Institute (STScI), 3700 San Martin Drive, Baltimore, MD 21218, USA}

\author[0000-0001-6052-3274]{Gabriel Caminha}
\affiliation{Max-Planck-Institut f\"ur Astrophysik, Karl-Schwarzschild-Str. 1, D-85748 Garching, Germany   (ORCID: 0000-0001-6052-3274)}

\author[0000-0001-8183-1460]{Karina Caputi}
\affiliation{Kapteyn Astronomical Institute, University of Groningen, P.O. Box 800, 9700AV Groningen, The Netherlands}
\affiliation{Cosmic Dawn Center (DAWN), Copenhagen, Denmark}









\begin{abstract}
   The first JWST data on the massive colliding cluster El Gordo confirm 23 known families of multiply lensed images and identify 8 new members of these families. Based on these families, which have been confirmed spectroscopically by MUSE, we derived an initial lens model. This model guided the identification of 37 additional families of multiply lensed galaxies in the  JWST data, among which 28 of them are entirely new system candidates and 9 are previously known. The initial lens model also helped determine geometric redshifts for the 37 additional systems. The geometric redshifts agree reasonably well with spectroscopic or photometric redshifts when those are available. The geometric redshifts enable two additional models that include all 60 families of multiply lensed galaxies spanning a redshift range $2 \lesssim z \lesssim 6$.  The derived distribution of dark matter confirms the double-peak configuration of the mass distribution of El Gordo found by earlier work with the southern and  northern clumps having similar masses. We confirm that El Gordo is the most massive known cluster at $z>0.8$ and has an estimated  virial mass close the maximum mass allowed by standard cosmological models.
   The  JWST images also reveal the presence of small-mass perturbers  that produce small lensing distortions. The smallest of these  is consistent with being a dwarf galaxy at $z=0.87$ and has an estimated mass of $3.8\times10^9$~\Msol, making it the smallest substructure found at $z>0.5$. The JWST images also show several candidate caustic-crossing events. One of them is detected at high significance at the expected position of the critical curve and is likely a red supergiant star at  $z=2.1878$. This would be the first red supergiant found at cosmological distances. The cluster lensing should magnify background objects at $z\ga6$, making more of them visible than in blank fields of similar size, but there appears to be a deficiency of such objects.

   
 \end{abstract}

%

\section{Introduction}
ACT-CL J0102$-$4915, known as El~Gordo and arguably the most famous galaxy cluster at redshift $z>0.8$,  has been observed with JWST/NIRCam as part of the PEARLS GTO project  \citet{Windhorst2022}. 
At $z=0.870$, \EG\ is a merging cluster with a double-peaked galaxy distribution \citep{Williamson2011,Menanteau2012}.
In X-rays, El Gordo exhibits a cometary structure clearly visible  in Chandra data. This suggests that El Gordo is being observed right after a collision of two subgroups \citep{Molnar2015,Zhang2015}, similar to the iconic Bullet Cluster. The presence of two radio relics ahead of and behind the X-ray cometary structure supports this interpretation \citep{Molnar2015,Lindner2014}. \cite{Ng2015} argued that El Gordo is in a return phase after first core passage. That is, the cluster is being observed after the phase of maximum separation, and  the two groups are moving back towards each other. This interpretation is, however, challenged by lens models based on strong lensing,  which place most of the mass in the SE group \citep{Zitrin2013,Cerny2018,Diego2020}. The massive nature of El Gordo is already demonstrated by its strong Sunyaev-Zeldovich (SZ) effect. Despite its relatively high redshift, \EG~shows up in Planck maps as a massive cluster with a signal-to-noise ratio of $\approx 13$ \citep{PSZ22016}. Some of the SZ signal may be due to increased pressure from the ongoing merger. 

\begin{deluxetable}{cCCl}
\tablecaption{El Gordo Mass Estimates}
\tablehead{
\colhead{Method\rlap{\tablenotemark{a}}}&
\colhead{Quantity\rlap{\tablenotemark{b}}}&
\colhead{Mass ($10^{15}$\,M$_{\odot}$)}&
\colhead{Reference}}
\startdata
SZ & M_{200m} & 1.89\pm0.15 & \cite{Williamson2011}\\
SZ & M_{200m} & 2.16\pm0.32 &
\cite{Menanteau2012}\\
SL & M_{200m} & 2.3 & \cite{Zitrin2013}\\
WL & M_{200c} & 3.13\pm0.56 & \cite{Jee2014}\\
SZ & M_{500c} & 1.07\pm0.05 & \cite{PSZ22016}\\
WL & M_{200c} & 1.11 & \cite{Schrabback2018}\\
SL & M(500~\rm kpc) & 1.1 & \cite{Cerny2018}\\
SL & M_{200c} & 1.08\pm0.70 & \cite{Diego2020}\\
WL & M_{200c} & 2.13 & \cite{Kim2021}\\
SL & M(1~\rm Mpc) & 1.84 & \cite{Caminha2022}\\
SL & M(500~\rm kpc) & 0.80\hbox{--}0.86 & This paper\\
SL & M(R_{\rm vir}) & 2.09\hbox{--}2.24 & This paper\\
\enddata
\tablenotetext{a}{SZ = Sunyaev-Zeldovich, SL = strong lensing, WL = weak lensing.}
\tablenotetext{b}{$M_{200m}$ is the mass within a sphere of radius $r_{200m}$, which is the radius above which the mass density drops below 200$\times$ the mean mass density of the Universe.  $M_{200c}$ is similar but measured within the radius where mass density drops below 200$\times$ the critical density, and $M_{500c}$ is the mass enclosed within the radius above which the density falls below 500 times the critical density. Other masses are measured within the specified physical radii. All $M_{200m}$ and $M_{200c}$ SL masses are obtained by extrapolating the profile obtained in the constrained region. }
\label{t:mass}
\end{deluxetable}

\EG's mass has been estimated  through a variety of techniques as shown in Table~\ref{t:mass}.
These methods agree that
El Gordo is the most massive cluster in its redshift range with an estimated mass 
above $10^{15}$~\Msol.
The most recent measurement \citep{Caminha2022} is notable because its strong-lensing model was based on a set of 23 spectroscopically confirmed systems. Prior to that work, all lens models had relied on photometric redshifts.  \\

At the upper end, the derived cluster masses are in tension with the standard $\Lambda$CDM model \citep[see for instance][]{Jee2014}, which finds a total mass exceeding  the maximum mass expected at this redshift at ${\lesssim} 2 \times 10^{15}~M_{\odot}$ \citep{Harrison2012}. A similar conclusion is reached from large N-body simulations. Using the very large 630 Gpc$^3$ N-body simulation Jubilee (based on a standard $\Lambda$CDM model), \cite[][their Fig.~5]{Watson2014} found that the most massive cluster at $z=0.9$ is expected to have $M_{178_m} \approx 1.7\times 10^{15}~M_{\odot}$. (\cite{Watson2014} defined the masses as $M_{178_m}$ rather than $M_{200_m}$ or $M_{200c}$. For an NFW profile, $M_{200_m} \approx 1.2M_{200c}$, and $M_{178_m}\approx 4\%$ times higher than $M_{200_m}$, hence this mass would correspond to  $M_{200_c} \approx 1.4\times 10^{15}M_{\odot}$.) 
The observed masses are mostly above this maximum mass but can become smaller if one considers  Eddington bias \citep{Waizmann2012}. 
Despite that, El Gordo is still uncomfortably close to the expected  $\Lambda$CDM limit.  El Gordo was found in a survey (the original ACT survey) that covered less than 2\% the area of the sky. The low likelihood of finding such an extreme object in such a small area raises the  tension with the $\Lambda$CDM model.
\cite{Asencio2021} used the same Jubilee simulation to argue that for El Gordo the tension is high.
On the other hand, after correcting for redshift and Eddington bias, \cite{Waizmann2012} found El Gordo not to be in tension with $\Lambda$CDM\null. However, their analysis assumed the survey area to be 3.7 times higher than the area where El Gordo was originally found. Any tension can be reduced if previous mass estimates are  too high. Additional mass estimates, based on alternative methods, can explore the uncertainty in the mass of the cluster. 

The X-ray emission exhibits an interesting offset between the peak of the X-ray emission and the position of the brightest cluster galaxy (BCG)\null. Contrary to what happens in the Bullet cluster, the X-ray peak seems to be ahead of the BCG\null. However, in the interpretation of \cite{Ng2015}, the BCG would be moving towards the second group, so the X-ray peak would be trailing the BCG\null. The returning-phase interpretation of El Gordo is challenged by dedicated N-body/hydrodynamical simulations reproducing most of the observations of El Gordo \citep{Molnar2015,Zhang2015}.
\cite{Molnar2018} demonstrated that the speed of the outgoing shocks can be very large (4000--5000~km s$^{-1}$) in a massive, merging cluster like El Gordo, therefore leaving the system before the first turnaround.

El Gordo is also unique in that it is a powerful lens at relatively high redshift.  One of the features that makes El Gordo an attractive target for lensing studies is the fact that for sources at high redshift, critical curves form at relatively large distances from the member galaxies. This is particularly true in the gap between the two clusters, where the critical curves are relatively undisturbed by nearby member galaxies. Having undisturbed critical curves is relevant to observe caustic crossing events of distant stars \citep{Kelly2018,Diego2018}, because in this case the maximum magnification can be larger than in situations where critical curves are affected by microlenses in member galaxies or  the intracluster medium. Caustic crossing events have been proposed as a technique useful to study Population~III stars and stellar-mass black-hole accretion disks  with JWST \citep{Windhorst2018}.

Because El Gordo is the highest-redshift known cluster with potentially  significant transverse motion---based on the X-ray
morphology and the two lensing mass centers discussed in this paper---it is an ideal target for JWST follow-up to search for caustic transits at $z\gg1$ and possibly for  caustic transits at $z>7$. For this reason, El Gordo was selected as a JWST GTO target. This paper uses the new JWST data to derive the mass distribution. We use our free-form lensing reconstruction code WSLAP+ \citep{Diego2005,Diego2007,Sendra2014,Diego2016,Diego2022}, which does not rely on assumptions about the distribution of dark matter. Our results offer an important cross-check with previous results because any disagreement between our free-form method and results obtained by previous parametric methods could signal potential systematic problems in one (or both) types of modeling. 

This paper is organized as follows. Section~\ref{sect_data} describes the data and simulations used in this work. Section~\ref{sect_constraints} describes the lensing constraints, and Section~\ref{sect_WSLAP} explains the lens modeling method and gives results for three models. 
The integrated mass and estimated virial masses for all three models are discussed in Section~\ref{sec_Mass} together with a comparison with earlier models in the literature.
Section~\ref{sect_perturbers} discusses small-mass perturbers found near a giant arc nicknamed ``La Flaca.'' 
Section~\ref{Sect_CaustCross} presents candidates to be compact, luminous objects near caustics and introduces "Quyllur," likely a red supergiant star that is being magnified by a factor of several thousand.   
Section~\ref{Sect_Discuss} discusses the key results, and Section~\ref{sect_concl} summarizes our conclusions. 
We adopt a standard flat $\Lambda$CDM cosmological model with $\Omega_m=0.3$ and $h=0.7$. At the redshift of the lens, this cosmological model implies that 1\arcsec\ corresponds to 7.8~kpc.

\section{JWST observations and ancillary data}\label{sect_data}

   \begin{figure*}
        \includegraphics[width=18cm]{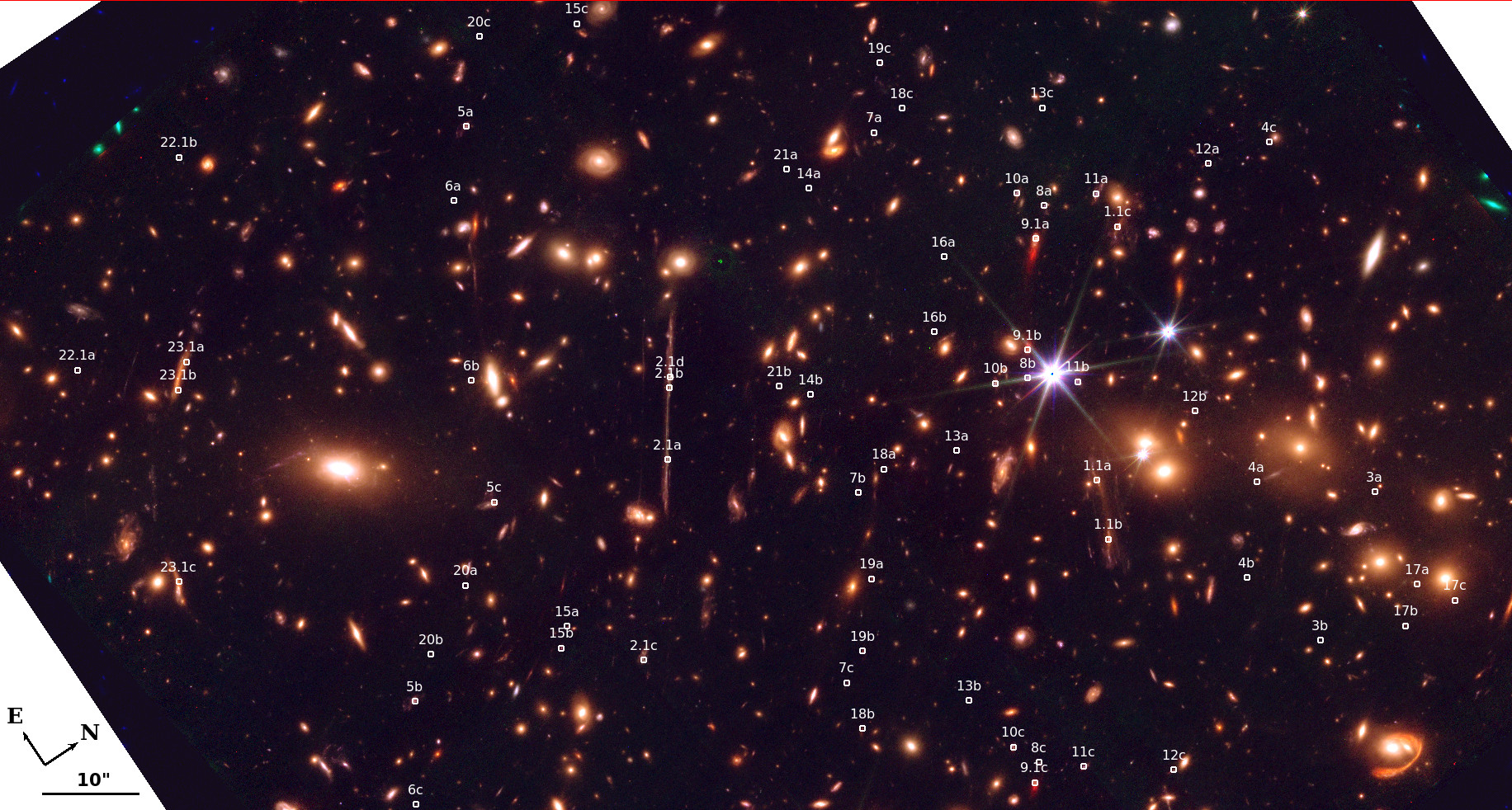}         
        \includegraphics[width=18cm]{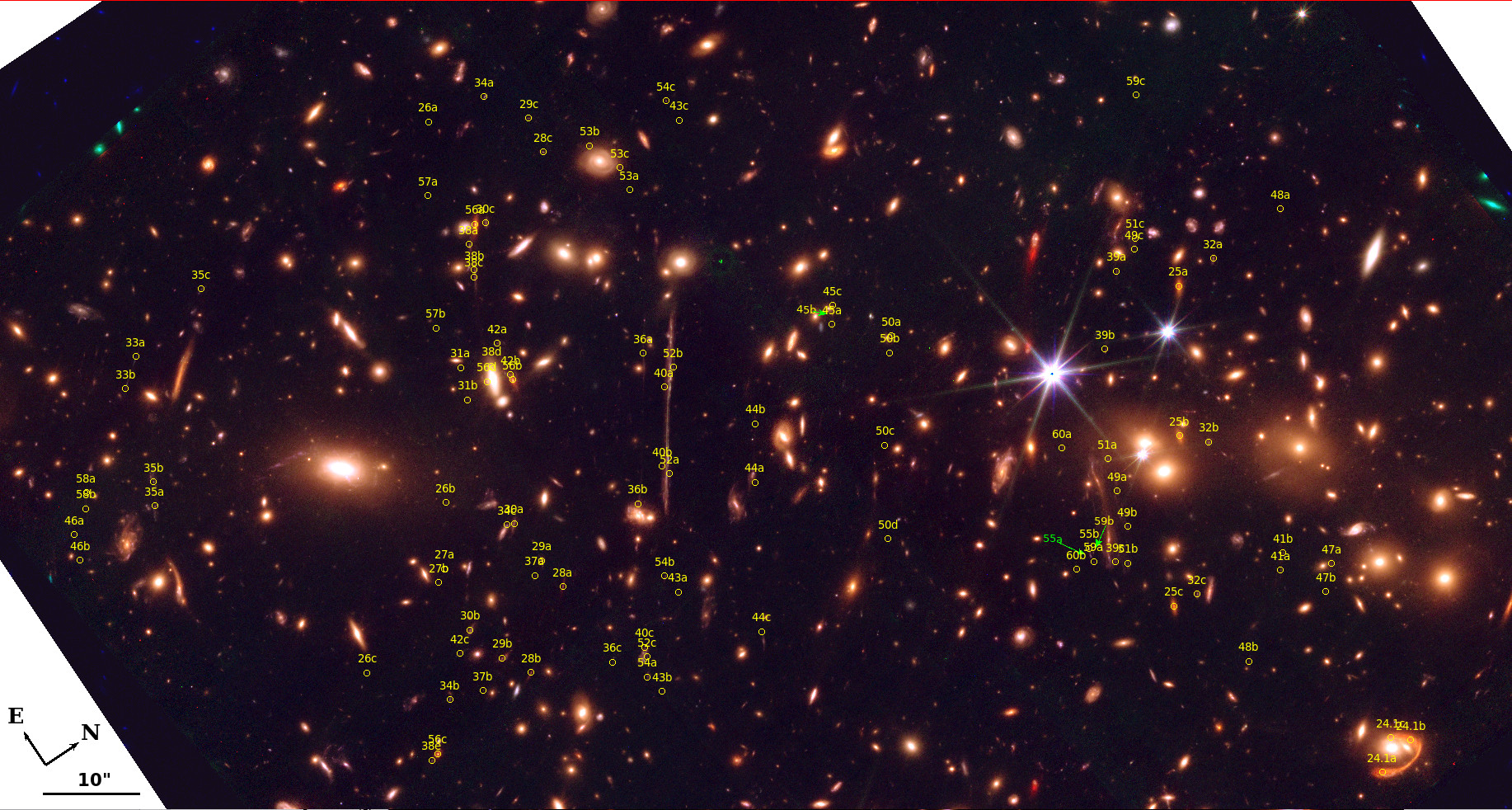}          
      \caption{Composite image of El Gordo obtained after combining HST and JWST bands (filters spanning from $\approx 0.5$ micron to $\approx 5$ micron). Top panel: white circles mark the systems confirmed spectroscopically by MUSE\null.  Bottom panel: yellow circles mark new system candidates identified with the new JWST data.
      In each label, numbers identify multiple images of the same background source, and letters identify the individual images. Image orientation and scale are indicated on the lower left.
              }
         \label{Fig_ElGordo}
   \end{figure*}


\subsection{JWST data}
El Gordo was observed by JWST on 2022 July 29 as part of the PEARLS Cycle~1 GTO program (pid \#1176, P.I. R. Windhorst).  Data were obtained in the IR filters F090W, F115W, F150W, F200W, F277W, F356W, F410M, and F444W with effective exposure times of 1889~s for F150W and F356W, 2104~s for F200W and F277W, and 2490~s for F090W, F115W, F410M, and F444W\null.
The 5$\sigma$ magnitude limit is $\approx$29.9 in the long-wavelength filters. The best spatial resolution was obtained in the F150W filter with a measured FWHM of 0\farcs063. Further details are  in the PEARLS overview paper \citep{Windhorst2022}. Figure~\ref{Fig_ElGordo} shows an overview of the field.

\begin{figure} 
      \includegraphics[width=8.5cm]{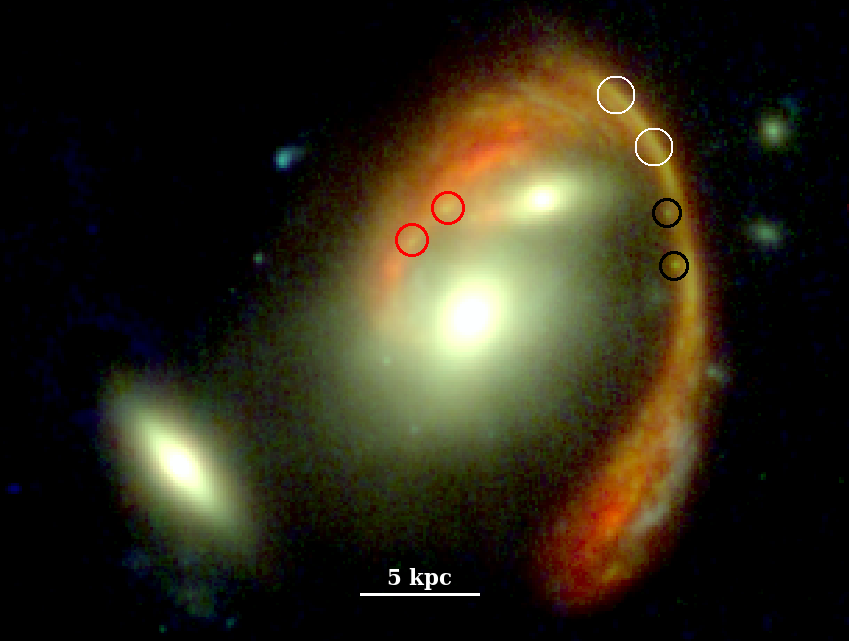}
      \caption{El Anzuelo galaxy in the northwestern part of the cluster. The galaxy has a photo-$z$ of $3.54$ \citep{Cheng2022}. The color-coded circles mark image pairs of three lensed sources.  Each pair identifies a critical point between the images.   
         }
         \label{Fig_AnzueloI}
\end{figure}

JWST reveals details of this cluster with unprecedented clarity.  Galaxies barely detected or not detected at all in HST data are easily detected by JWST, and many are much brighter at wavelengths longer than 2~\micron\ than at the longest HST wavelength (1.6~\micron).  Figure~\ref{Fig_AnzueloI} shows one such galaxy, which is heavily distorted by cluster-galaxy lensing into a fishhook shape. We will refer to this galaxy as "El Anzuelo" (Spanish for fishhook). 
Pairs of multiply lensed features are easily identified in the image. Image pairs constrain the position of critical curves, which are expected to pass  between the images. The red nucleus of the extended background galaxy is lensed into at least three images clearly visible in the reddest JWST bands. The three images of the nucleus coincide with the position of ALMA sources EG-SMG~2 and EG-SMG~4 \citep{Cheng2022}, the latter containing two images barely resolved by ALMA\null. Previous HST data did not show El Anzuelo   as multiply lensed because this galaxy is red and therefore  faint at wavelengths shorter than 1.6~$\mu$m. HST data do, however, identify a multiply-lensed blue feature.

\begin{figure*} 
   \includegraphics[width=\linewidth]{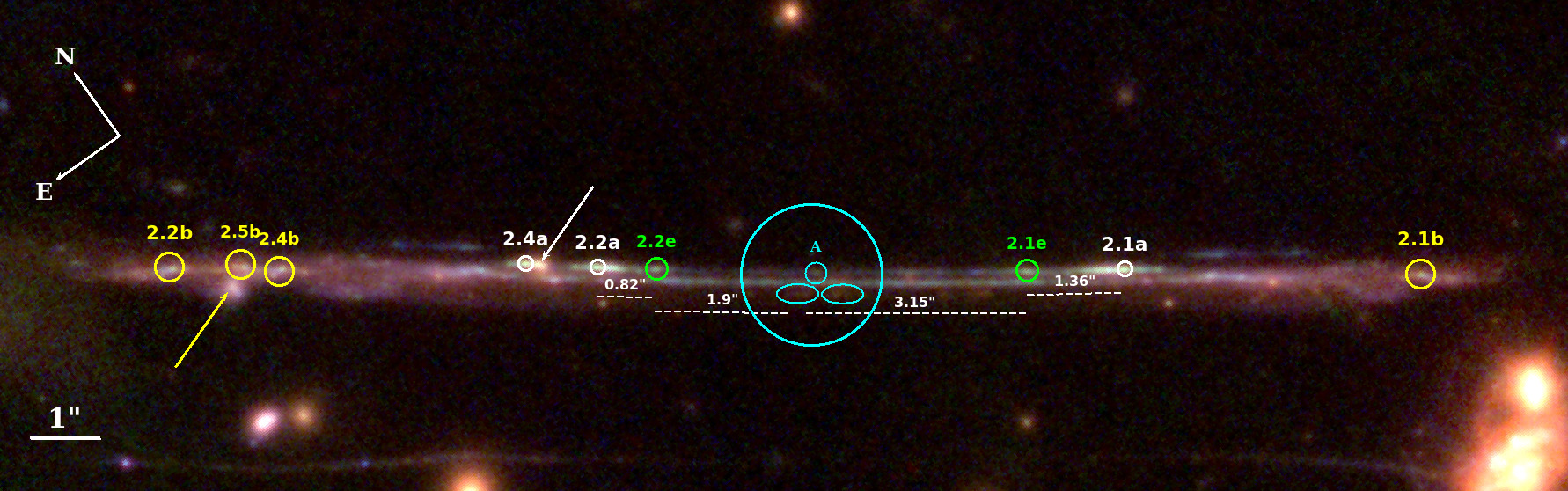}
      \caption{Perturbers in the giant arc La Flaca.  Scale and orientation are shown on the image. Color-coded circles mark multiple images of three background sources. La Flaca is extremely thin and one of the longest known arcs. It is located between the two main mass clumps in El Gordo. Small-mass perturbers create additional multiple images. These perturbers are marked with arrows.  The approximate position of the critical curve can be estimated by extrapolating the ratio of distances between knots 2.1a--2.1e and knots 2.2a--2.2e. The point where one expects the critical curve is marked with the cyan circle having 1\arcsec\ radius. The two cyan ellipses with in this circle mark a possible double counterimage. If these two objects are counterimages of each other, the critical curve would be very close to the middle point between the two. The small cyan circle labeled A marks a possible caustic-crossing object.  
         }
         \label{Fig_LaFlaca}
\end{figure*}

Another  example illustrating the JWST data quality is shown in Figure~\ref{Fig_LaFlaca}, which shows the giant arc already observed in pre-JWST data. We refer to this galaxy as "La Flaca" ("the thin one" in Spanish). Unlike El Anzuelo, La Flaca is easily visible in HST data. 
The new JWST data show with greater detail the multiply-lensed features within this arc. In particular, the nucleus of the arc appears three times (2.1a, 2.2a, and 2.4a)  instead of the expected two times. The extra image (2.4a, already seen in HST data) is in fact a double image (unresolved) that appears as a consequence of a small perturber marked with a white arrow in the figure. Another perturber splits counterimage 2.2b into three images (2.2b, 2.4b, and 2.5b).  Section~\ref{sect_perturbers} discusses these perturbers in greater detail. Unlike El Anzuelo, La Flaca does not show any clear pair of counterimages that can be used to constrain the position of the critical curve. However, it is still possible to get an approximate location based on simple arguments. One can use the ratio of distances between knots 2.1a--2.1e and 2.2a--2.2e to derive the relative magnification between counterimages (Figure~\ref{Fig_LaFlaca}). This ratio is 1.66. If we assume that this magnification ratio is maintained between knots 2.1e and 2.2e, the critical curve is approximately $5\farcs05/(1+1.66)=1\farcs9$ southwest from knot 2.2e as shown in Figure~\ref{Fig_LaFlaca}. (The distance between knots 2.1e and 2.2e (5\farcs05.) This is an approximation because the magnification ratio may change slightly as one approaches the critical curve, but it should be  a good approximation. A pair of features near the expected position of the critical curve,  could be counterimages of each other bracketing the critical curve. 

\subsection{HST data}
In addition to JWST data, we used public HST imaging data from programs GO 12755 (P.I. J.~Hughes), GO 12477 (P.I. F.~High), and GO 14096 (P.I. D.~Coe). These ACS and WFC3/IR observations include data in 10 filters spanning wavelengths $\sim$0.4--1.6$\mu$m. The Reionization Lensing Cluster Survey \citep[RELICS;][]{Coe2019} delivered reduced images combining data from all of these HST programs, including their own (14096). The RELICS data release also includes galaxy catalogs  with photometry and photometric redshifts.
 
The HST data add valuable information, especially at wavelengths shorter than 0.8~\micron. The HST data also cover a wider field of view than JWST\null. This is important for lens modeling because some member galaxies are outside the field of view of JWST but  within the HST field.  An especially important one is a massive member galaxy (at $\rm RA=15.752995$, $\rm Dec=-49.281048$ and spectroscopically confirmed at $z=0.8749$ by \citeauthor{Caminha2022}) that falls on the edge of the JWST image.

The color images in this paper include both HST and JWST data.  A particular combination, which we call RGB6, offers a good compromise between low noise levels, sensitivity, and color range. $RGB$ colors are defined by
\begin{align}
\nonumber
R&=\rm 0.3\times F277W({JWST})
+0.35\times F356W({JWST})\\
&\rm +0.15\times F410M({JWST})
+0.2\times F444W({JWST}),\\ 
G&=\rm 0.45\times F150W({JWST})
+0.55\times F200W({JWST}), \\ 
\nonumber
B&=\rm 0.1\times F606W({HST})
+0.1\times F625W({HST})\\
&\rm +0.3\times F090({JWST})
+0.4\times F115({JWST})\quad.
\end{align}
Unless otherwise noted, all color images shown in this work are based on RGB6. 


\subsection{X-ray data}
\begin{figure} 
      \includegraphics[width=8.5cm]{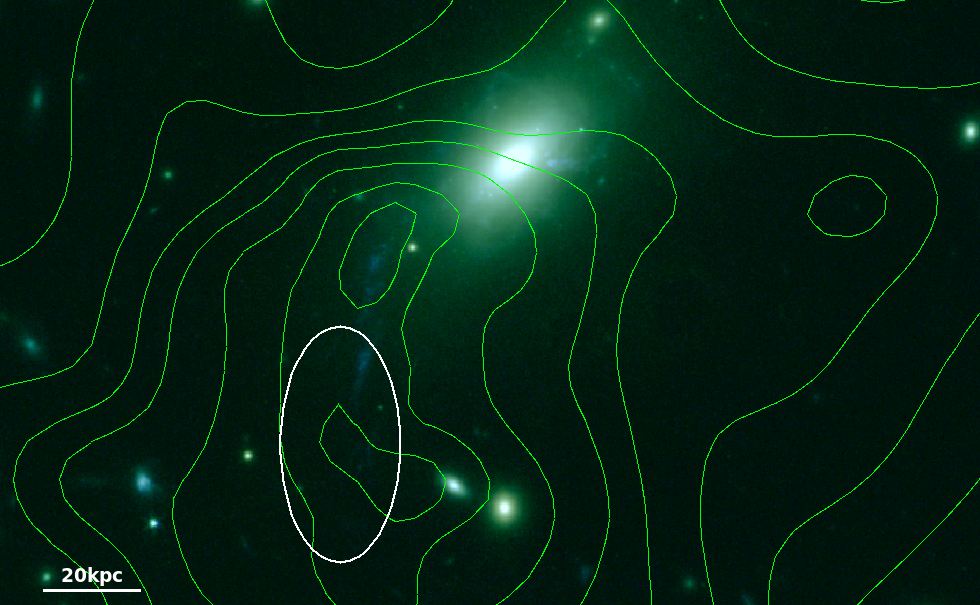}
      \caption{JWST image of the central BCG with smoothed X-ray contours shown in green. Scale bar shows physical distance corresponding to 2\farcs56.  The white ellipse marks the position and beam shape (ATCA at 2.1 GHz) of the radio point source U7  \citep{Lindner2014}.  Blue knots in and above the circle agree in position with [\ion{O}{2}] emission seen by MUSE\null.}
         \label{Fig_BCG_JWST}
\end{figure}

Chandra data from the ACIS instrument acquired in 2011--2012 (ObsID 12258, 14022, and 14023, PI.. J.~Hughes) total $\approx$350~ks. The X-ray data were smoothed using the code {\sc asmooth} \citep{Ebeling2006}. 
Figure~\ref{Fig_BCG_JWST} shows the BCG galaxy with overlaid contours of the smoothed X-ray data. Gas falling into the BCG forms a filamentary structure that coincides with the peak X-ray emission, suggesting  a powerful cooling flow.
The infalling gas has strong emission in [\ion{O}{2}] and correlates spatially with the peak of the X-ray emission. [\ion{O}{2}] line in MUSE data reveals a velocity structure in the infalling gas. 
The intense X-ray emission combined with the infalling velocity structure suggest this filament is the optical counterpart of a powerful cooling flow, where large amounts of material fall towards the BCG\null. The strong [\ion{O}{2}] emission in the filament can be understood as the plasma being chemically enriched by feedback process from the BCG itself or from material being stripped away from other member galaxies.   \cite{Lindner2014} found a radio point source ``U7'' at one of the extremes of the filament. No obvious counterpart for U7 can be identified in JWST images, but when one accounts for the beam of the radio data, U7 coincides with the peak of the X-ray emission and the infalling [\ion{O}{2}]-emitting gas. U7's radio emission may be synchrotron emission arising as the result of rapid deceleration of electrons where the  cooling plasma encounters  denser material surrounding the BCG.

\section{Lensing constraints}\label{sect_constraints}
In the strong lensing regime, lens models are primarily constrained by the observed positions of multiply-lensed galaxies. The first step is recognizing which observed objects constitute an ``image family'' of a single background source.  Once a suite of image families from sources at different redshifts are available, the lens model is optimized by searching for a solution that focuses each family of images into a single position at the corresponding redshift. The redshift information is ideally obtained from spectroscopy. When this is not available,  photometric redshifts can be used instead, or the redshift can be treated as a free parameter. 

With the new data from JWST, some candidate image families found in earlier work \citep{Zitrin2013,Cerny2018,Diego2020} are now confirmed, thanks to the improved resolution, depth, and color information. The improvement in spatial resolution and depth also unveils new  candidates. We visually inspected the JWST images and compiled a new list of  candidate families. Spectral information from MUSE and  redshifts derived by \cite{Caminha2022} confirmed some of these candidates as real multiple-image systems. 

Using MUSE data, \cite{Caminha2022} confirmed 23 families of lensed galaxies, 12 of which were previously known and 11 of which were new systems. System~8 of \cite{Caminha2022} is a redefinition of System~6 of \cite{Diego2020} but with two counterimages incorrectly identified. Of the 23 \citeauthor{Caminha2022}  systems, 12 had only two images visible. The third image is presumably too faint to be identified in the HST images or show up in the MUSE spectra. Thanks to JWST's increased depth and spectral range, we found candidates for the third image for 8 out of these 12 systems. For the remaining 4, the counterimage is expected to be too faint to be visible even in the current JWST data. 
The list of all available constraints with spectroscopic redshift confirmation is given in Table~\ref{tab_arcs}.  

In addition to the classic constraints given by the observed positions of strongly lensed images, we also used the position of critical points that act as anchors for the critical curve. Critical points are positions that a critical curve is known to pass through. These can be identified as symmetry points in arcs that are formed by the merging of two images. Several of these merging arcs can be found in the JWST data. One needs to know the redshift of the arc in order to use the critical point as a valid constraint, and therefore only critical points of galaxies that have spectroscopic redshift can be used. There are three such critical points in systems 1, 2 and 23. Their positions are listed at the end of Table~\ref{tab_arcs}. 

\begin{figure}
  \centering
   \includegraphics[width=8cm]{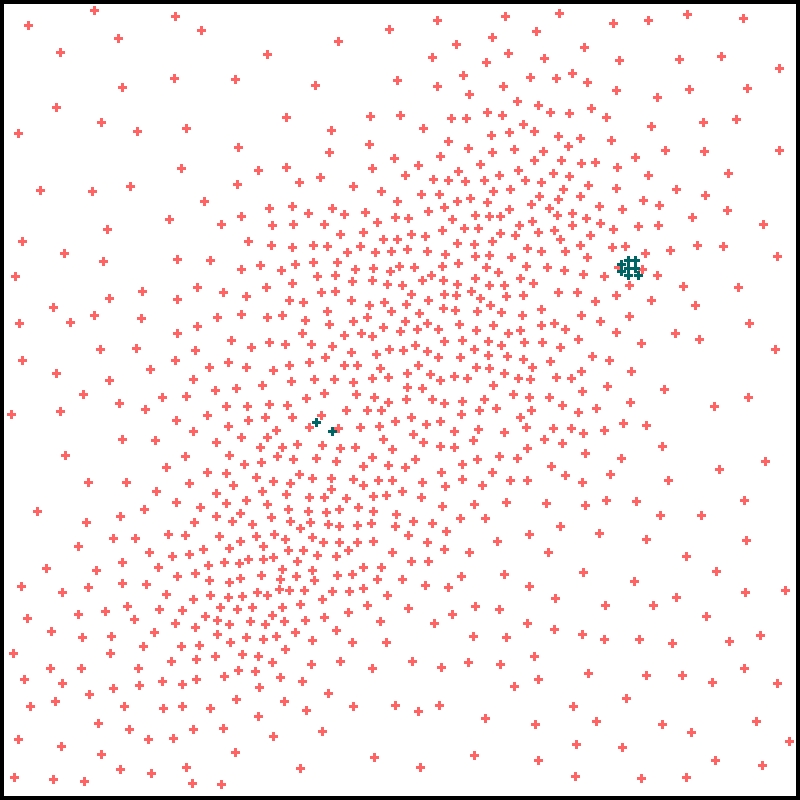}
     \caption{Multiresolution adaptive grid. The number density of grid points (shown as 891 red crosses) traces a smooth version of the solution in the spectroscopic lens model, increasing the resolution in the densest regions.  Ten additional grid points are shown as small black crosses.  Two of them mark the perturbers in the La Flaca arc with scales of 0\farcs12 and 0\farcs24, and eight more are in the inner region of the Anzuelo arc. Each one of these 8 grid points has a scale of 1\farcs8. 
              }
         \label{Fig_Grid}
   \end{figure}

\section{Free-form lens modeling of El Gordo}\label{sect_WSLAP}
\subsection{Modeling method}
The mass reconstruction is based on our method WSLAP+\null. Previous papers \citep{Diego2005,Diego2007,Sendra2014,Diego2016} give details. As a brief summary, the lens equation is
\begin{equation} 
\beta = \theta - \alpha(\theta,\Sigma(\theta)), 
\label{eq_lens} 
\end{equation} 
where $\theta$ is the observed position of the lensed source, $\alpha$ is the deflection angle, $\Sigma(\theta)$ is the surface mass-density of the lens (El Gordo cluster in our case) at  position $\theta$, and $\beta$ is the true position of the background source. Both the strong-lensing and weak-lensing data can be expressed in terms of derivatives of the lensing potential $\psi$:\footnote{Note however, that through observations one measures the reduced shear, $\gamma_r=\gamma/(1-\kappa)$ where $\kappa$ is the convergence.}
\begin{equation}
\psi(\theta) = \frac{4 G D_{l}D_{ls}}{c^2 D_{s}} \int  \Sigma(\theta')\ln(|\theta - \theta'|)\,d^2\theta' \quad, 
\label{eq_psi} 
\end{equation}
where $D_l$, $D_s$, and $D_{ls}$ are the angular-diameter distances to the lens, to the source, and from the lens to the source, respectively. The unknowns of the lensing problem are the surface mass density and the positions of the background sources in the source plane. 

As shown by \cite{Diego2005,Diego2007}, the lensing constraints can be expressed as a system of linear equations  
\begin{equation}
\vectTheta = \matrGamma \vectX\quad, 
\label{eq_lens_system} 
\end{equation} 
where the observables (positions of strongly lensed galaxies, positions of critical points,  and weak lensing if available) are contained in the array $\vectTheta$, and the unknown surface mass density, re-scaling factors for the different layers discussed above, and true source positions are in the array $\vectX$. The matrix $\matrGamma$ is known and given by the position of the grid points and positions of the constraints. In our case, $\vectTheta$ contains the positions listed in Table~\ref{tab_arcs}  (lensed galaxies and critical points). (No weak-lensing constraints were used in this work.) Details on how the critical points are added to the system of linear equations are given by \cite{Diego2022}.
A solution, or lens model, is found  by minimizing a quadratic function derived from the system of linear equation with the constraint  $\vectX>0$. That is, all masses and source positions (referred to a corner of the field of view) must be positive. 


The model surface mass-density for El Gordo was described by a combination of two components: 
i) a soft (or diffuse) component and 
ii) a compact component that accounts for the mass associated with the individual halos (galaxies) in the cluster. 
The diffuse component was computed as a set of mass centers at pre-defined positions. Each mass center was assumed to represent a surface mass-distribution defined by a Gaussian with a pre-defined full width at half maximum (FWHM)\null. 
The algorithm then optimized the mass of each mass center to best fit the constraints. 

Mass distributions other than Gaussians could have been used, but Gaussian functions provide a good compromise between the desired compactness and smoothness  of the basis function. A Gaussian basis offers several advantages, including a fast analytical computation of the integrated mass for a given radius, a smooth and nearly constant amplitude between overlapping Gaussians (with equal amplitudes) located at the right distances, and  orthogonality between relatively distant Gaussians that help reduce unwanted correlations.  (\citealt{Diego2007}  discussed alternative basis functions including polynomial, isothermal, and power laws.)

For the compact component, we adopted  the light distribution around the brightest member galaxies in the cluster.  For each location, we assigned a mass proportional to the galaxy surface-brightness at that location with the mass-to-light ratio ($M/L$) being solved for as part of the optimization process. We used the surface brightness in the F160W HST image because its  wide field of view includes some galaxies not covered during by JWST\null.  
The compact component was divided into four independent layers with independent $M/L$ because $M/L$ can differ for different galaxy types.  The first layer contains only the main BCG\null. The second layer contains the three  galaxies acting on El Anzuelo. Layer~3 contains all remaining members in the cluster, and Layer~4 contains a small group of galaxies in the foreground at $z=0.63$ \citep[][]{Caminha2022}. As discussed by \citeauthor{Caminha2022}, although this is a small group, it has a non-negligible contribution to the lensing.

\begin{figure}
  \centering
   \includegraphics[width=9cm]{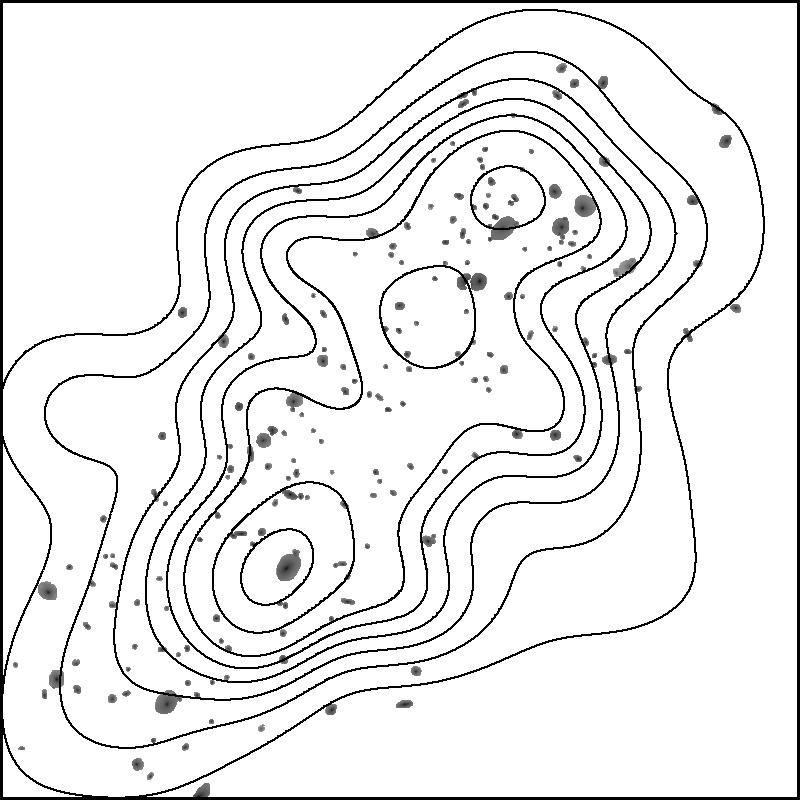}
     \caption{Spectroscopic lens model. Contours represent the mass distribution of the smooth component of the lens model, i.e., dark matter, diffuse baryons such as stars from the ICL, and X-ray emitting plasma. The galaxies used to describe the compact contribution to the lens model are shown in gray. 
     North is up and east is left. The image is 2.5 arcminutes across. 
              }
         \label{Fig_Driver_vs_Galaxies}
   \end{figure}

\begin{figure}
  \centering
   \includegraphics[width=9cm]{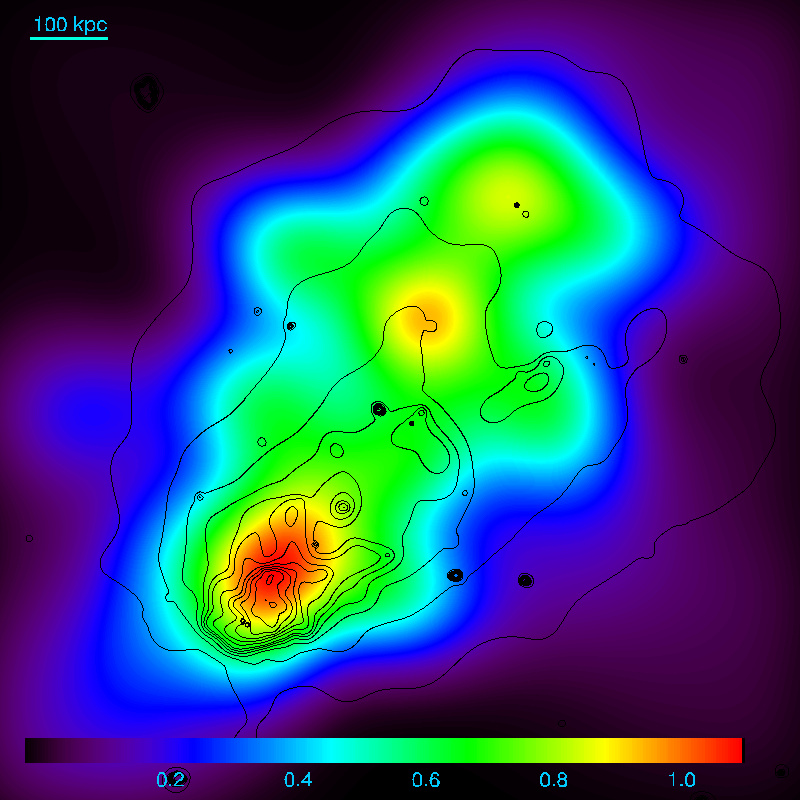}
     \caption{Diffuse mass component from the spectroscopic model vs X-ray emission from Chandra.  Color  shows the convergence at $z=3$ as indicated by the color bar. Black contours show the X-ray emission from Chandra. The image is oriented north up, east to the left, and a scale bar is shown at top left. 
              }
         \label{Fig_Driver_vs_Xray}
   \end{figure}

\subsection{Spectroscopic lens model}\label{sec_Driver}
We derived a first model (the ``spectroscopic model'') using only 23 families 
that have spectroscopic redshifts. These are identified as rank~A  in Table~\ref{tab_arcs}. 
The soft-component mass-centers were defined on a uniform 32$\times$32 grid.  This is equivalent to adopting a flat prior for the mass distribution.
The grid spacing was 4.6875\arcsec, and the FWHM of each Gaussian mass center was taken to be 2\farcs4.

The solution for the spectroscopic model is shown in Figure~\ref{Fig_Driver_vs_Galaxies}. The distribution of the soft component (mostly dark matter) correlates well with that of the member galaxies. A clear peak in the soft component appears at the position of the  BCG  in the southeast part of the cluster. In contrast, the northwest clump has a less-concentrated peak in the soft component, in agreement with previous results. 

Figure~\ref{Fig_Driver_vs_Xray}  compares the  soft mass component  with the X-ray emission from Chandra. The mass in the southeast clump peaks close to the position of the apparent cooling flow (Section~\ref{sect_data}). 
On larger scales, the distribution of X-rays shows a cometary structure similar to the Bullet cluster. The X-ray morphology  suggests that the southern clump is moving in the southwest direction, and therefore that the cluster has already had one encounter. The northern clump is more irregular and may have been affected by the dynamics of the collision. 
There is an excess of mass east of the local X-ray peak. The south clump is much more massive than the northern one but also more elongated along the apparent direction of motion.

\subsection{Geometric redshifts from the spectroscopic lens model}\label{Sect_SLz}


The spectroscopic lens model 
relies only on confirmed systems, and therefore it gives a robust solution unaffected by redshift uncertainties of the background sources. 
This model can be used to confirm new system candidates and predict their redshifts. These redshifts are known as geometric redshifts (or geo-$z$) because they are the redshifts at which the multiple images from a given family focus at a single point. That is, the redshift corresponds to the focal plane of the lens for that particular family of images.  

The spectroscopic model gives geometric redshifts for all system candidates listed in Table~\ref{tab_arcs}. The probability of a system to be at redshift $z$ is
\begin{equation}
    P(z) = \exp(-V(z)/[2\sigma^2)]\quad,
    \label{eq_Pz}
\end{equation}

where $V(z)$ is the variance between the multiple positions of a given system projected on the source plane at redshift~$z$. The projection was calculated from the deflection field of the spectroscopic model (computed at $z=3$) which was re-scaled for each system to the corresponding redshift. The dispersion, $\sigma$, in the expression above was fixed to 0\farcs18 (about 3 pixels in the LW channels). This is a reasonable choice for well-constrained systems, resulting in relatively narrow distributions for the redshift. Systems that are well reproduced by the spectroscopic model result small $V(z)$ near the optimal redshift, which in turn results in maximum values of $P(z)$ close to 1. Systems that are poorly reproduced by the spectroscopic model have larger values of $V(z)$ at its minimum, which reduces the maximum value  of $P(z)$.
A low maximum probability for $P(z)$ does not necessarily mean that the system is a bad candidate. This can simply be the result of the spectroscopic model not being well constrained in that part of the lens plane.  
Systems at high redshift tend to have broader probabilities because  the deflection field varies slowly with redshift for $z\ga4$.

\begin{figure} 
      \includegraphics[width=8.5cm]{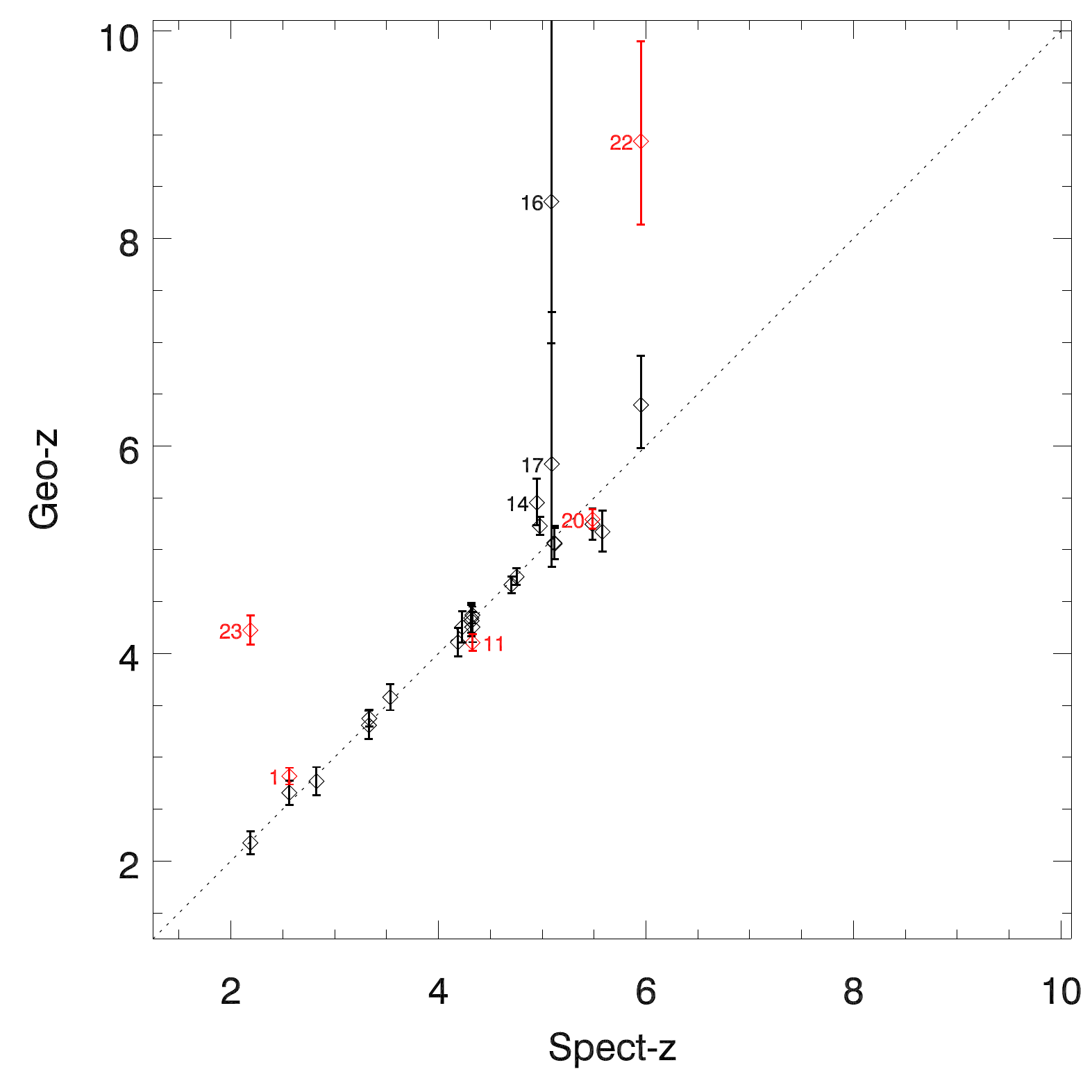}
      \caption{Geometric redshifts compared with the input spectroscopic redshifts. This plot shows a sanity test where the input redshifts are properly recovered by the lens model. Families with discrepant redshifts are labeled. Black points show results of the spectroscopic model, and red points show results of a bootstrap analysis where a new model was derived without that particular system and that model used to predict the system's redshift. 
         }
         \label{Fig_SLz}
\end{figure}

As a sanity check, we used the spectroscopic model to derive geometric redshifts of the systems used to build the model. The result is shown in Figure~\ref{Fig_SLz}. As expected, most systems were recovered with a redshift close to the true redshift and with small uncertainties.   Systems~22 and~23 lie at the edge of the region that has lensing constraints. Their uncertainties are larger, and the model may not be reliable in those areas.  The only other clear outlier is system 16, which has only two counterimages (as does System~14, which agrees reasonably well). System~17 has a large uncertainty because its counterimages are close to each other and near member galaxies that are not individually constrained. 


\begin{figure} 
      \includegraphics[width=8.5cm]{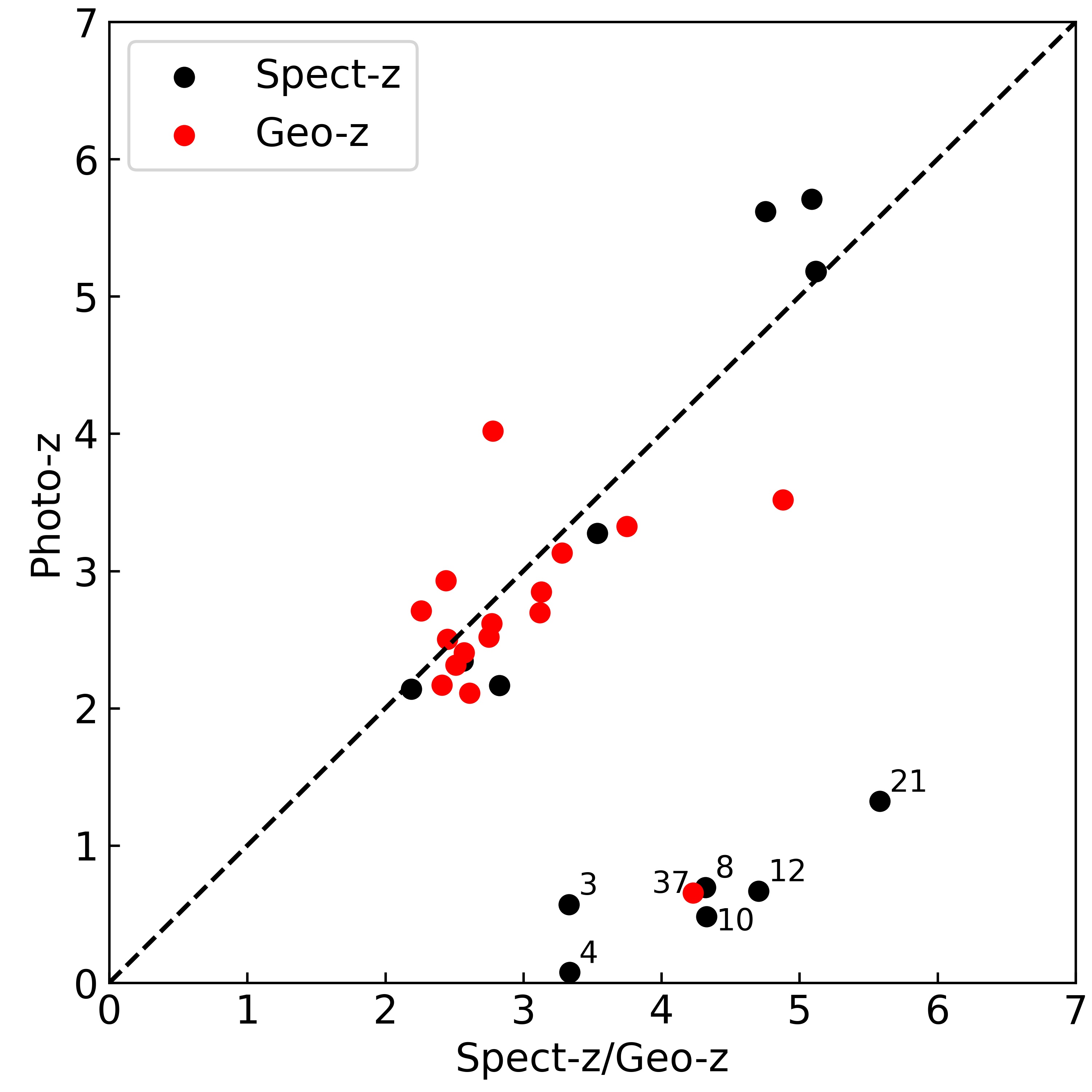} 
      \caption{Comparison of photometric redshifts with other redshifts. Black points represent spectroscopic redshifts, and red points represent geometric redshifts, both of which are shown on the $x$ axis. Dashed line shows equality, and discrepant systems are labeled.  Only systems having consistent photo-$z$ estimates for at least two counterimages are shown.
         }
         \label{Fig_GeoZ_vs_PhotoZ}
\end{figure}

A better way to check the model is to bootstrap the lensed systems. We removed one system  at a time and derived a new lens model, then used that model to predict the redshift of the system that was removed.  In general the spectroscopic model  still recovered the redshift of these systems with good accuracy. The exception are Systems~22 and~23, which are at the outskirts of the constrained region. Therefore removing one of these systems results in an accentuated degradation of the spectroscopic model. Where several spectroscopic systems exist, redshifts are well recovered. 
In earlier work, a similar  bootstrap  confirmed the excellent performance of geo-$z$ estimates of well-calibrated lens models  \citep{Chan2017,Chan2020}. 

Geometric redshifts are a useful alternative to photometric redshifts, especially for high-redshift candidates for which photometry may be poor or nonexistent (e.g., dropouts). Also, photometric redshifts may be unreliable for dusty galaxies at $z\approx4$ because their Balmer break may be misinterpreted as Ly-$\alpha$, placing them at much higher redshifts \citep{Naidu2022,Zavala2022,Harikane2022}.

To provide photometric redshifts, we used
LePhare \citep{Arnouts1999,Arnouts2011}. The setup was the same as used with NIRCam data by \cite{Adams2022}. In brief, sources were selected in  the  F444W  image using SExtractor \citep{Bertin1996}. Its dual-image mode was then used to derive photometry ({\sc mag\_auto}) in all other filters. Galaxy templates were the BC03 set \citep{Bruzual2003} with 57 ages, a Chabrier IMF  \citep{Chabrier2000}, constant or exponential star formation histories, solar or 20\% solar metallicity, dust extinction in the range of $0<E(B-V)<1.5$ \citep{Calzetti2000}, the IGM treatment from \cite{Madau1995}, and  $0\le z< 25$.   Figure~\ref{Fig_GeoZ_vs_PhotoZ} shows the results. 
In general, photo-$z$ agrees with spec-$z$ and geo-$z$. The outlier points in the horizontal branch are mostly biased low photo-$z$ estimates although  three of the seven labeled outliers in Figure~\ref{Fig_GeoZ_vs_PhotoZ} (Systems~4, 8, and~21) have secondary solutions much closer to the  spectroscopic redshift. In a wider sample of galaxies with spectroscopic redshifts (not necessarily multiply lensed), $\approx$86\% have LePhare redshifts with an error $\delta z/(1+z)<0.15$.


\begin{figure}
  \centering
   \includegraphics[width=9cm]{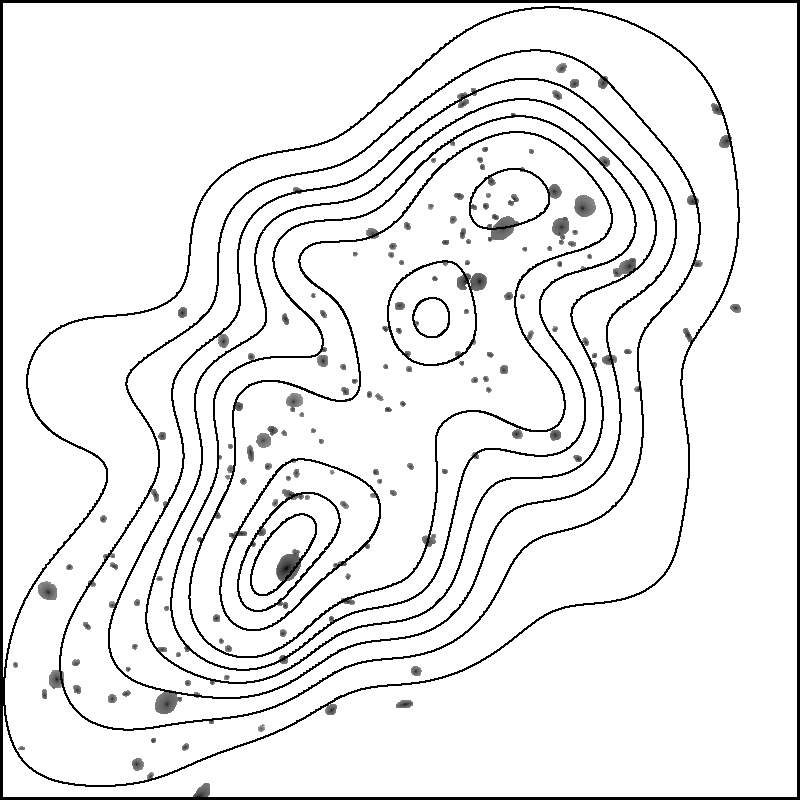}
     \caption{Diffuse mass distribution and  galaxy locations for the Full-r model. Contours represent the smooth component (dark matter and diffuse baryons such as stars from the ICL and X-ray-emitting plasma). The galaxies used to describe the compact contribution to the lens model are shown in gray. 
              }
         \label{Fig_Full_vs_Galaxies_FULLr}
   \end{figure}

A general problem with photometric redshifts based on  NIRCam  observations alone is that they can give multiple redshift solutions. Some of this degeneracy can be rectified by the inclusion of other data such as from HST\null. Also, residual systematics in zero-point calibrations may still be affecting the photo-$z$ estimates with JWST \citep{Boyer2022}. 
RELICS photometric redshifts (based on BPZ: \citealt{Benitez2000}), given in Table~\ref{tab_arcs}), offer examples of the improvement that can be obtained. In particular, adding HST data gives fewer small-phot-$z$ outliers. For example, for System~8 with $z_{\rm spec}=4.3175$,  RELICS predicts $z_{\rm phot}\approx 4.3$ while JWST photometry alone predicts $z_{\rm phot}\approx 0.7$. System~10 at $z_{\rm spec}=4.3275$ is consistent with the RELICS $z_{\rm phot}\approx$ 4.2--4.6, and in better agreement than JWST-based $z_{\rm phot}\approx 0.7$. This is almost identical to System~12 ($z_{\rm spec}=4.7042$) where HST  $z_{\rm phot}\approx$ 4.6--5.3 while JWST  $z_{\rm phot}\approx 0.7$.   
System 21 at $z_{\rm spec}=5.5811$ is predicted to be at $z_{\rm phot}\approx 6$ according to HST while JWST predicts $z_{\rm phot}\approx1.3$. 
\cite{Frye2022} will provide additional photometric redshifts that combine HST and JWST photometry. 
The important lesson from Figures~\ref{Fig_SLz} and~\ref{Fig_GeoZ_vs_PhotoZ} is that geo-$z$ can be reliable estimates of the redshift. 

\begin{figure}
  \centering
   \includegraphics[width=9cm]{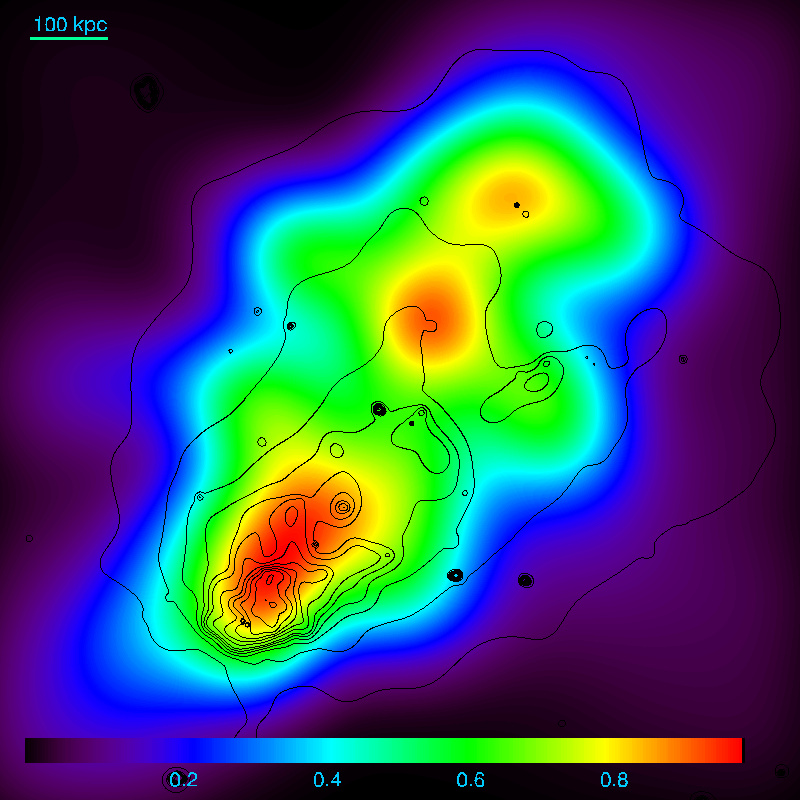}
     \caption{Diffuse mass component from the Full-r  model vs X-ray emission from Chandra. Color  shows the convergence at $z=3$ as indicated by the color bar. Black contours show the X-ray emission from Chandra. The image is oriented north up, east to the left, and a scale bar is shown at top left. 
              }
         \label{Fig_Full_vs_Xray_FULLr}
   \end{figure}

\subsection{Lens model with geometric redshifts}\label{sec_Full}

One of the factors determining the uncertainty of a lens model is the possible error in the redshifts. 
This is particularly problematic for high-redshift galaxies that are visible only in the reddest JWST bands, mimicking galaxies at even higher redshifts as recently studied by \cite{Naidu2022,Zavala2022,Harikane2022}. 
The spectroscopic model is a robust solution because it relies on systems with spectroscopic redshifts, but the model is limited by the redshifts available.

In order to improve the resolution of the spectroscopic model, we need to include systems without spectroscopic redshifts.  To do so without adding bias from inaccurate photometric redshifts, we relied on the geometric redshifts. This adds 37 additional system candidates labeled  in Figure~\ref{Fig_ElGordo} (bottom).  The ``Full-r model'' maintains the $32\times32$ regular grid used in the spectroscopic redshift model and simply increases the number of constraints by using all 60  systems with rank~A or~B listed in Table~\ref{tab_arcs}. 

\begin{figure}
  \centering
   \includegraphics[width=9cm]{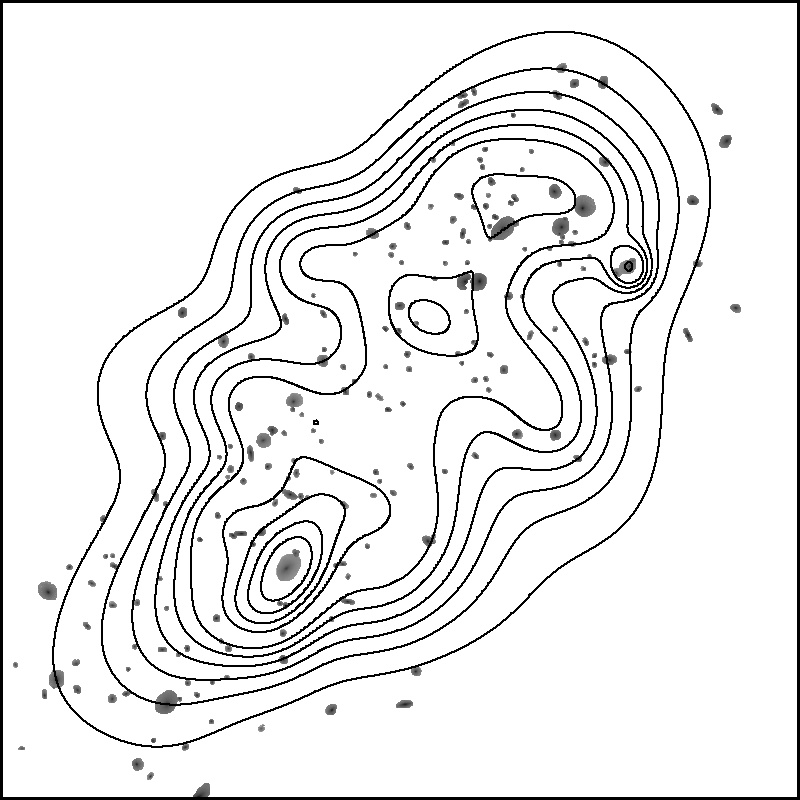}
     \caption{Mass vs galaxies for the Full-a model. Contours represent the mass distribution of smooth component (dark matter and diffuse baryons such as stars from the ICL and X-ray emitting plasma). The galaxies used to describe the compact contribution to the lens model are shown in gray . 
              }
         \label{Fig_Full_vs_Galaxies_FULLa}
   \end{figure}

\begin{figure}
  \centering
   \includegraphics[width=9cm]{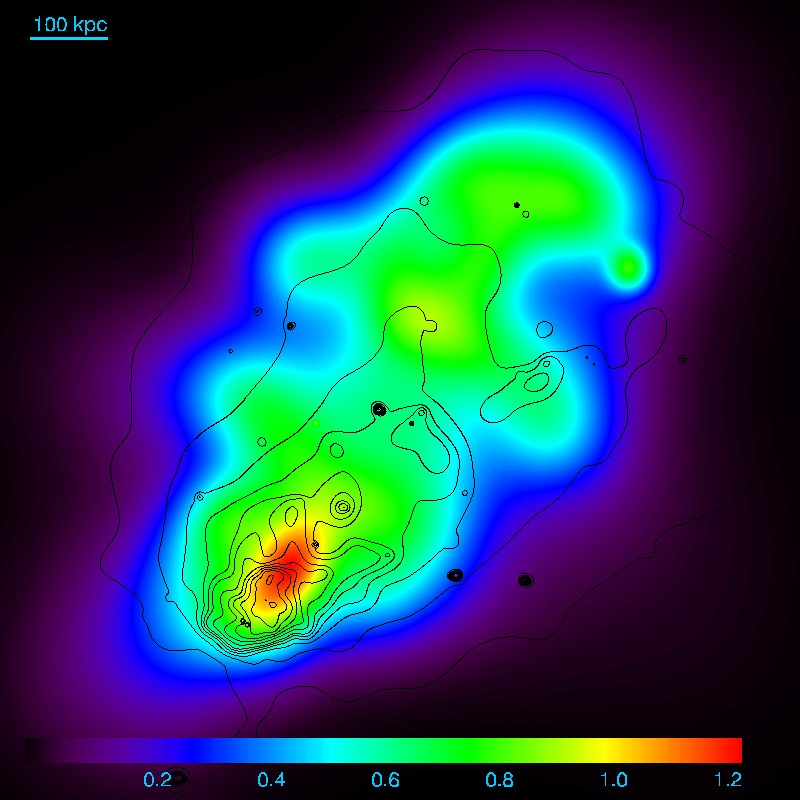}
     \caption{Mass vs X-ray for the model Full-a.  The color plot shows the convergence (at z=3) for the spectroscopic model. Only the dark matter component of the lens model is shown. X-ray emission from Chandra are shown as black contours. Note the excess of mass near the X-ray local peak at the east. The south clump is now clearly more massive than the northern one but also more elongated along the direction of motion.
              }
         \label{Fig_Full_vs_Xray_FULLa}
   \end{figure}

The Full-r model  solution  is shown in Figure~\ref{Fig_Full_vs_Galaxies_FULLr}. As in the spectroscopic model, there is a good correlation between the smooth and compact components. The new constraints reveal more detail in the distribution of the smooth component. The clump in the northwest shows a more irregular distribution than in the spectroscopic model. In between the two clumps, we see excess mass going in the east--west direction. This excess mass correlates well with the wings of the X-ray emission. Simulations of the cluster collision \citep{Molnar2015} show similar wings forming in the direction perpendicular to the axis of the collision provided the impact parameter is $\le$100~kpc and infalling velocities are $\ga$2500~km\,s$^{-1}$). The smooth component traces mostly the dark matter and the X-ray-emitting plasma, and the excess mass in the direction perpendicular to the collision axis could be partially due to the plasma. 
We also see an elongation of the smooth component in both clumps, especially the southern clump.  The peak of the smooth component correlates well  with the peak in the X-ray emission (Figure~\ref{Fig_Full_vs_Xray_FULLr}) although there is a small offset. This is partially due to the possible cooling flow that is the brightest feature in X-rays. Finally, the amplitudes of both mass peaks are slightly smaller in the Full-r model than in the spectroscopic model.

\begin{figure*} 
   \includegraphics[width=18cm]{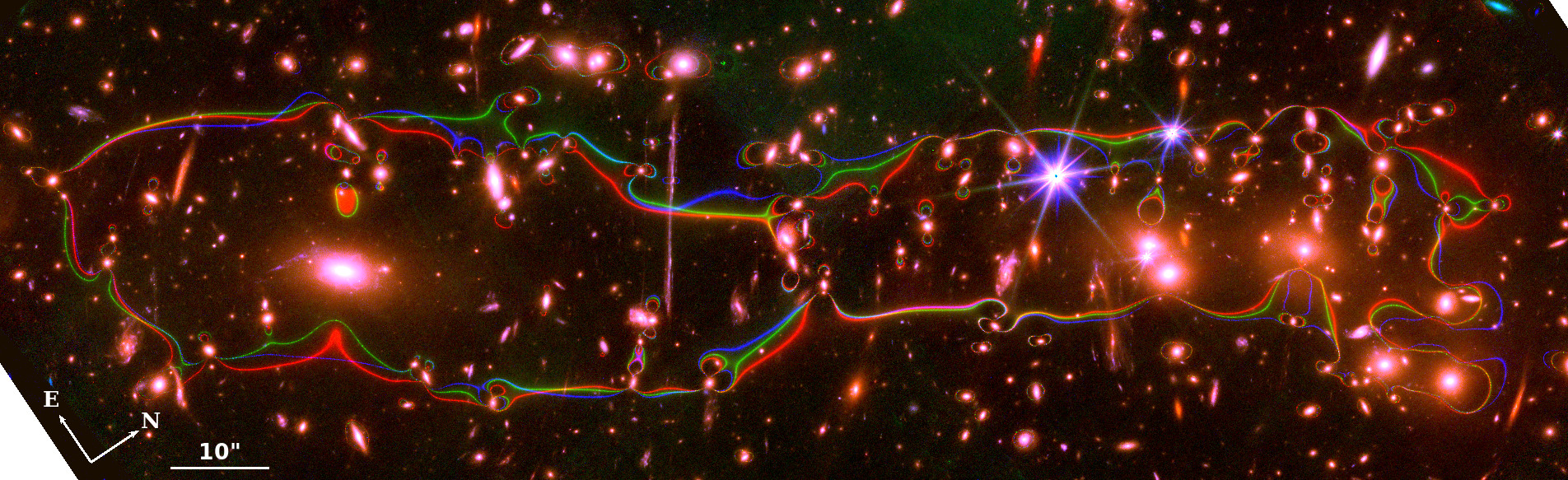}
      \caption{Comparison of the critical curves for the three models. The critical curves are computed at $z=2.8254$. The red curve corresponds to the spectroscopic model, the green curve is for model Full-r, and the blue curve is fr model Full-a. The color image is made from the combination of HST+JWST filters RGB6 discussed in section \ref{sect_data}. 
      }
         \label{Fig_ThreeCC}
\end{figure*}

\subsection{Lens model with adaptive grid}

To refine the mass distribution, we built a third model called ``Full-a.'' It has the same constraints as the Full-r model but placing the soft-component mass-centers on an irregular grid.  This provides increased resolution in regions with more mass.
The grid was built from a Monte Carlo realization of a smoothed version of the mass distribution of the spectroscopic model, and 
instead of fixed sizes of 2\farcs4, the 891 main mass centers have widths proportional to the average separation from their neighbors. 
This grid is shown in Figure~\ref{Fig_Grid}. 
It can be interpreted as using the prior that the mass distribution is the one indicated by the spectroscopic model, but of course the model optimization allows  the derived mass distribution to deviate from the prior.
The  grid also includes two Gaussians with widths 0\farcs12 and 0\farcs24  at the position of two perturbers near the arc nicknamed La Flaca (Figure~\ref{Fig_perturbersI}) and  eight additional Gaussians with FWHM 1\farcs8 around the group lensing the galaxy El Anzuelo. 
 
The Full-a model reveals even more details as shown in Figures~\ref{Fig_Full_vs_Galaxies_FULLa} and~\ref{Fig_Full_vs_Xray_FULLa}. The mass distribution has a more concentrated peak around the BCG in the SE clump, while the NW clump is less centrally concentrated though about the same mass. Neither the Full-a model nor any of the others shows excess mass at the position of the [\ion{O}{2}] filament. This suggests that the contribution of the dense filament to the projected mass must be relatively small.

\section{El Gordo mass estimate}\label{sec_Mass}
\begin{figure} 
   \includegraphics[width=9cm]{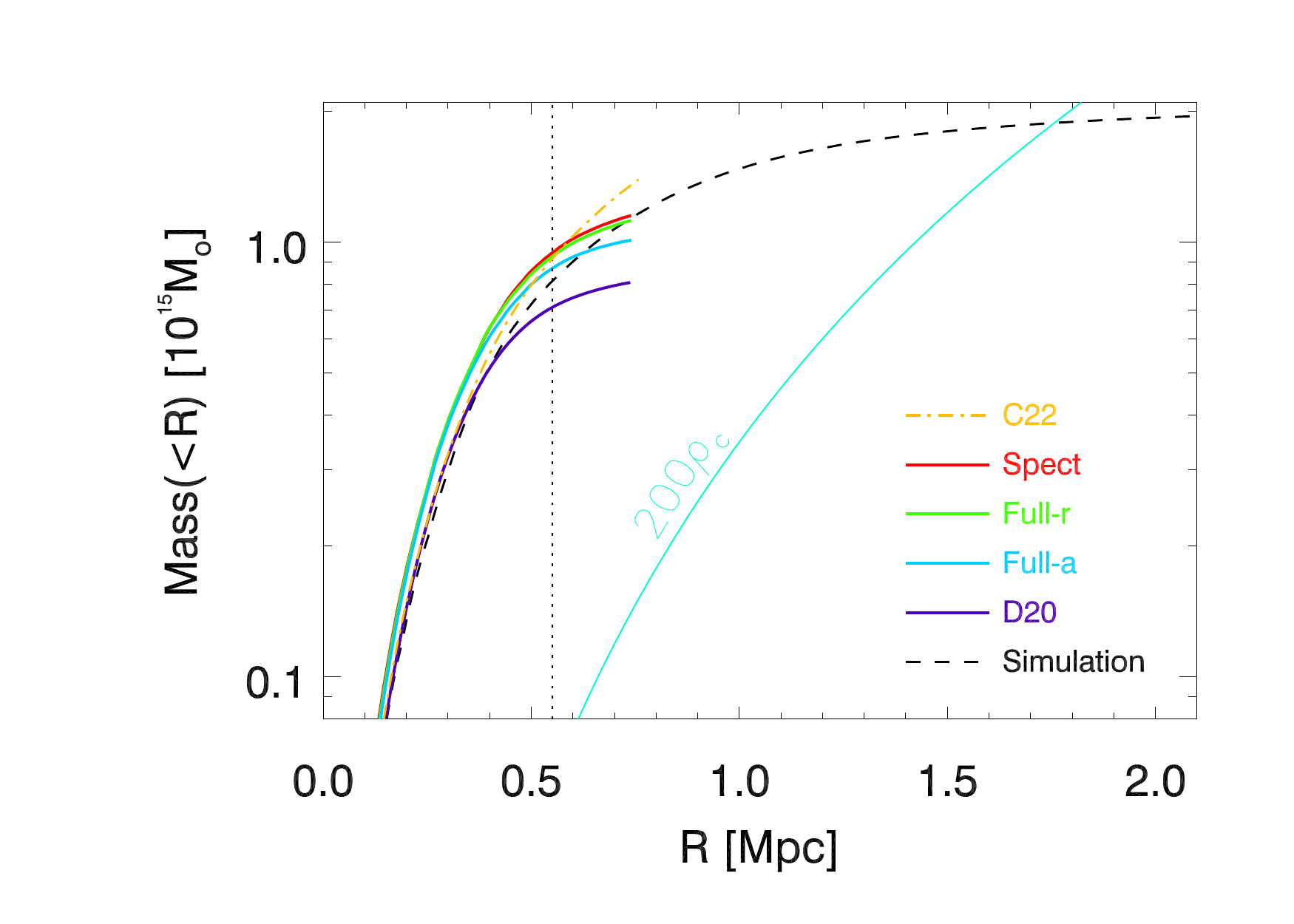}
      \caption{Total El Gordo mass as a function of radius. The red curve shows values for the spectroscopic model, the green curve for the Full-r model, and the cyan curve for the Full-a model. The purple line shows the mass obtained derived from the RELICS data  using the same algorithm but different constraints \citep{Diego2020}. The orange dot-dashed line shows the recent model of \cite{Caminha2022}. The black dashed line shows values from a simulation \citep{Molnar2015}. All mass curves are based on a common center.  The light turquoise curve shows the total mass enclosed in a sphere with constant density equal to the critical density. The vertical dotted line marks the maximum radius (70\arcsec) at which lensing constraints exist. The  integrated masses will be biased low beyond this point.
         }
         \label{Fig_TotalMass}
\end{figure}

\begin{deluxetable}{lC}
\tablecaption{El Gordo Clump Mass Estimates}
\label{t:gmass}
\tablehead{
\colhead{Model}&
\colhead{$M(\rm 300~kpc)$}}
\startdata
&{10^{14}~\Msol}\\
SE clump:\\
\cite{Caminha2022}& 2.29$^a$\\
Spectroscopic & 3.46\\
Full-r & 3.43\\
Full-a & 3.29\\
\tableline
NW clump:\\
\cite{Caminha2022}& 2.19$^a$\\
Spectroscopic & 3.85\\
Full-r & 3.79\\
Full-a & 3.67\\
\enddata
\tablenotetext{a}{Masses in \cite{Caminha2022} exclude the contribution from member galaxies}
\end{deluxetable}

All mass models show a clear double-peaked distribution. The southern group is centered near the BCG, and the northern group is centered near a luminous galaxy at $\rm RA=15\fdg7210573$, $\rm Dec=-49\fdg2528437$. Masses of each clump for the various models are given in Table~\ref{t:gmass}. The table includes masses from the parametric model of \cite{Caminha2022} for comparison.
The \citeauthor{Caminha2022} masses are lower because they exclude the mass of  member galaxies, while we quote total masses. Otherwise all masses agree reasonably well.

\begin{figure*} 
   \includegraphics[width=18.0cm]{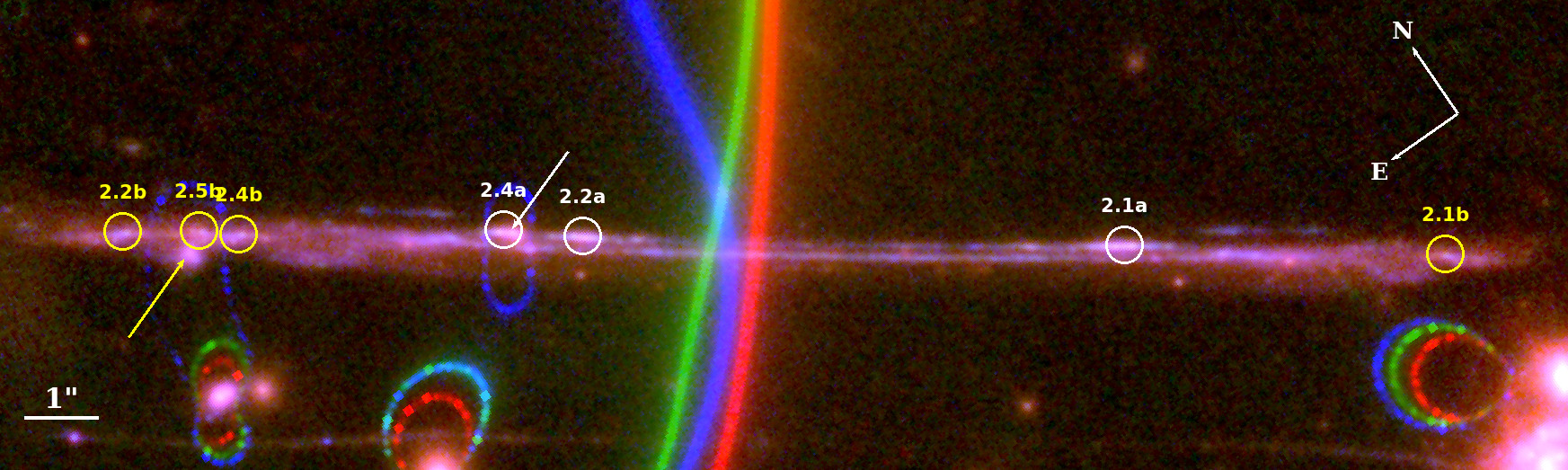}
      \caption{Small scale deflectors. The giant, thin arc nicknamed La Flaca runs horizontally across the image.  The heavy red, green, and blue curves are the critical curves of El Gordo at the redshift of the arc ($z_{\rm spec}=2.8254$) for the spectroscopic, Full-r and Full-a models, respectively. The Full-a model (but not the other models) includes  small deflectors at the two positions marked by the white and yellow arrows. Dotted blue curves show the critical curves of the deflectors. The deflector marked with a yellow arrow has a MUSE spectrum with $z= 0.7842$.  With a Gaussian FWHM of 0\farcs24, its mass is $2.7\times10^{10}$~\Msol. The deflector marked with a white arrow is barely detected, and we have assumed it is at El Gordo's redshift. With $\rm FWHM = 0\farcs12$, its mass is  $3.8\times10^{9}$~\Msol. 
         }
         \label{Fig_perturbersI}
\end{figure*}

The virial mass in El Gordo has been used in earlier work to study possible tension with the standard cosmological model, which predicts the most massive cluster at this redshift should have a virial mass $M_{200c}\lesssim 2\times 10^{15}$~\Msol\ \citep{Harrison2012,Watson2014}. 
Given the similarity in mass of the two clumps, the cluster center of mass should be near the midpoint between the two groups, and we adopted a galaxy at  $\rm RA=15.7313639$, $\rm Dec=-49\fdg2605609$.  
Figure~\ref{Fig_TotalMass} shows the  integrated masses as a function of radius for the models derived here and some others. At 500~kpc, the masses for the models described here  are between 8.0 and $8.6\times10^{14}$~\Msol. 
These are higher than the \cite{Diego2020} model based on RELICS HST data (M(500 kpc)$=6.6\times10^{14}$~\Msol) but in reasonable agreement given the different data sets.
The \citet{Caminha2022} model used exactly the same constraints as our spectroscopic model but  a completely different approach.  
The agreement within the constrained region is excellent, and   $M({500~\rm kpc})=7.98\times 10^{14}$~\Msol, also in good agreement with our estimates. 
Finally, the N-body simulation of \cite{Molnar2018} agrees very well with the lensing models up to ${R}\approx 500$~kpc, the limit of the strong-lensing constraints, and    $M(500~{\rm kpc})=7.20\times 10^{14}$~\Msol, also in good agreement.  Above 500~kpc, one expects El Gordo's mass profile to continue as in the simulation. The virial radius is 1.75~Mpc, and the simulation gives  $M(R_{\rm vir})=1.88\times 10^{15}$~\Msol.  At 500~kpc, the lens models give masses 11--19\% higher than the N-body model, and one might correct the N-body mass higher by the same factor. This is at least close to, and probably above, the cosmological limit. A more robust estimate of the virial mass could be obtained by combining the strong lensing constraints with weak lensing measurements extending beyond the virial radius.


\section{Small scale substructures in La Flaca}\label{sect_perturbers}
As discussed in Section~\ref{sect_data} and shown in Figure~\ref{Fig_LaFlaca}, the giant arc La Flaca contains two small scale perturbers. We can use the observed lensing distortions to infer their  masses as
WSLAP+ optimizes the mass in all grid points.
Figure~\ref{Fig_perturbersI} shows how the critical curve of the larger deflector winds around knots 2.2b, 2.5b, and 2.4b to reproduce the observed triple image. The curve of the smaller deflector  passes between knots 2.2a and 2.4a. We expect 2.4a to be a double image, but if so, it is unresolved in JWST images. The critical curve is close to the middle point in 2.4a but is still off by a small fraction of an arcsecond, suggesting there is still room for a small improvement in modeling this double knot.  The masses inferred by the lens model are consistent with small galaxies. In particular, the smaller 2.4a perturber has mass  consistent with being a dwarf galaxy if it is at El Gordo's redshift. As far as we know, this would be the smallest mass measured at $z>0.5$. The effective mass of a perturber scales as its mass times the macromodel magnification, and new JWST images near critical curves  perturbed by small halos will allow measuring masses for even smaller substructures, potentially constraining models of dark matter, as  \cite{Diego2022} recently did. 



\section{Caustic-crossing candidates}\label{Sect_CaustCross}

 Among the greatest achievements by HST in the last few years was its discovery of extremely magnified stars at high redshift.  Such objects were first predicted by \cite{MiraldaEscude1991}. The discovery of Icarus \citep{Kelly2018} marks the starting point of this new field, which allows  study not only of individual stars at $z>1$ but also compact star clusters  \citep{Dai2021}. The discovery also  opens the door to novel studies of dark matter structures \citep{Diego2018,Oguri2018,Venumadhav2017, Dai2018, Dai2020, Dai2020s1226,Diego2022,Meena2022}. After the discovery of Icarus, other examples followed \citep{Chen2019,Kaurov2019,Diego2022,Chen2022}, culminating with the recent discovery of Earendel at $z\approx 6.2$ \citep{Welch2022a,Welch2022b}. 
 
\begin{figure} 
    \includegraphics[width=8.5cm]{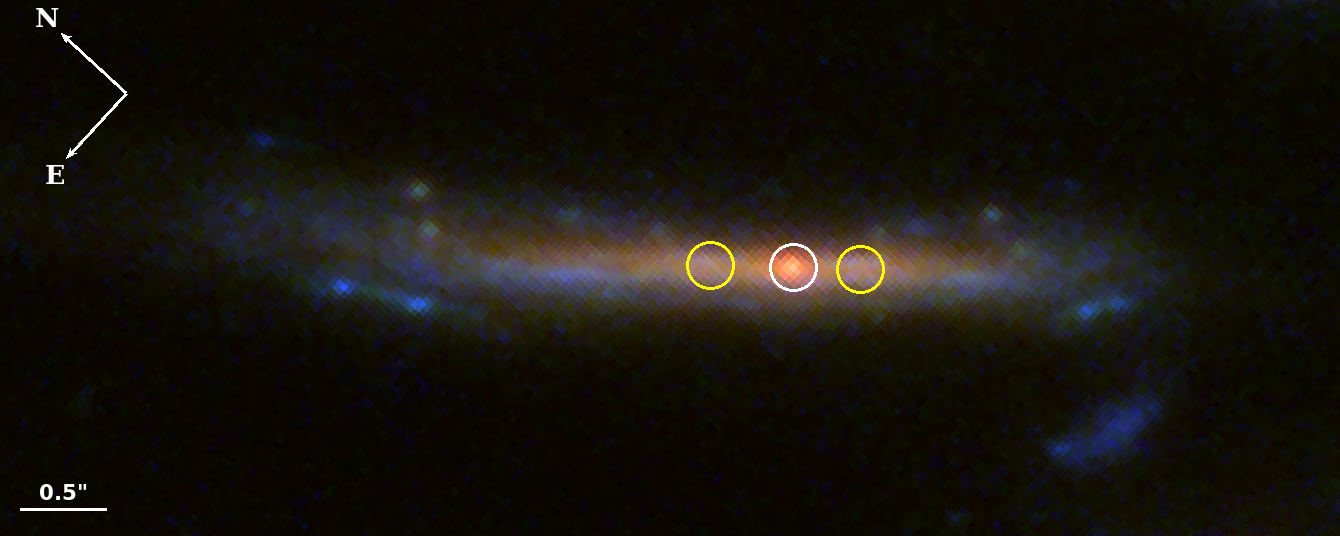}
    \includegraphics[width=8.5cm]{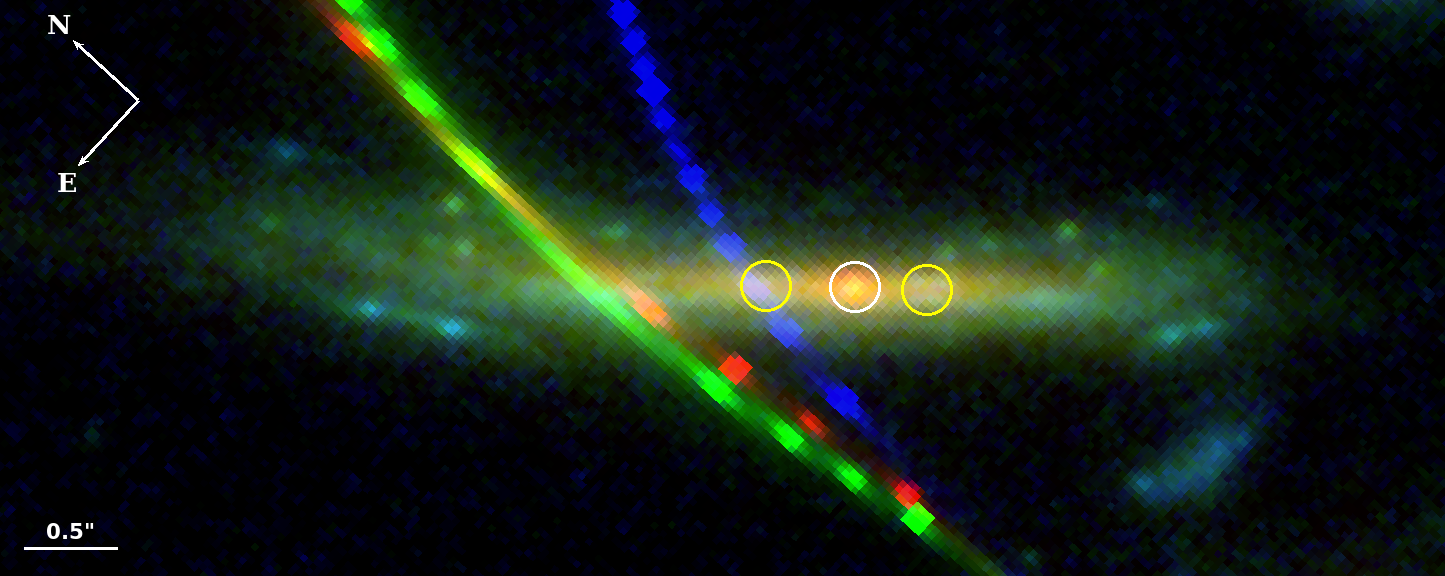}
      \caption{Possible caustic crossing event in system 23 at $z=2.1878$.
      {\bf Top:} Color image  combining  F115W, F200W, and F356W as blue, green, and red, respectively.
      The outer, yellow circles mark the positions of 
      23.4a and 23.4b,
      two counterimages of a source that brackets the position of the critical curve. The white, central circle marks a bright source (called ``Quyllur'') that lies approximately at the midpoint. {\bf Bottom: }
      color image with the RGB6 combination.
  Red, green, and blue curves show the critical curves for the spectroscopic, Full-r, and Full-a models respectively.  The two panels have the same scale and orientation as indicated at the left.
         }
         \label{Fig_Star5}
\end{figure}

JWST is expected to see further than HST and possibly detect individual stars close to the beginning of cosmic reionization \citep{Windhorst2018}. The first public data from JWST already show candidate lensed stars at cosmological distances \citep{Pascale2022}. A search for similar candidates in El Gordo found a clear example in a strongly lensed arc  shown in Figure~\ref{Fig_Star5}.  Given its proximity to the midpoint between the pair of knots 
23.4a and 23.4b (yellow circles in Figure~\ref{Fig_Star5}), this source (marked with a  white circle in Figure~\ref{Fig_Star5}) is likely right on  the critical curve and therefore highly magnified. Because images near a critical curve always come in pairs, the lack of counterimage and unresolved nature of the source implies that the image pair must have a counterimage separation no more than approximately 1~pixel or 30~mas. This situation is similar to the cases of Godzilla \citep{Diego2022} and Earendel \citep{Welch2022a,Welch2022b}, where those sources must also form unresolved pairs of counterimages.

Several alternative possibilities are discussed below, but the most likely explanation for the source is a single, highly magnified, red supergiant star. We nickname it ``Quyllur,'' which is the Quechua term for star.\footnote{Pronounced Koijur in English. Quechua was spoken by the Incas before the arrival of  Europeans, and it is still spoken by millions of people in different parts of South America. Similar to the disappearance of the Incan empire, Quyllur must have vanished eons ago. But similarly to the Quechua language that is still alive, Quyllur's light still traverses the Universe}.

\begin{figure} 
    \includegraphics[width=8.5cm]{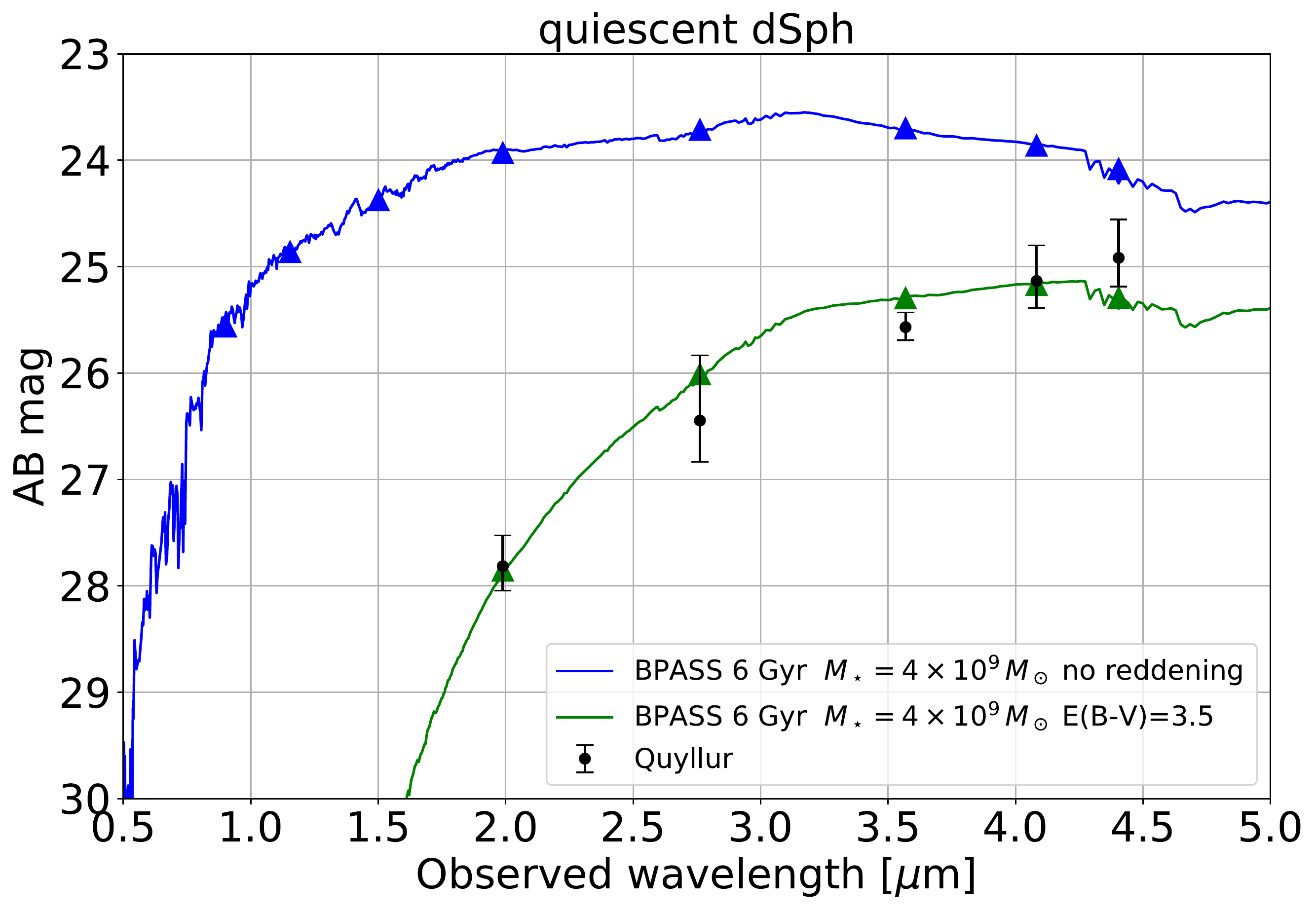}
    \includegraphics[width=8.5cm]{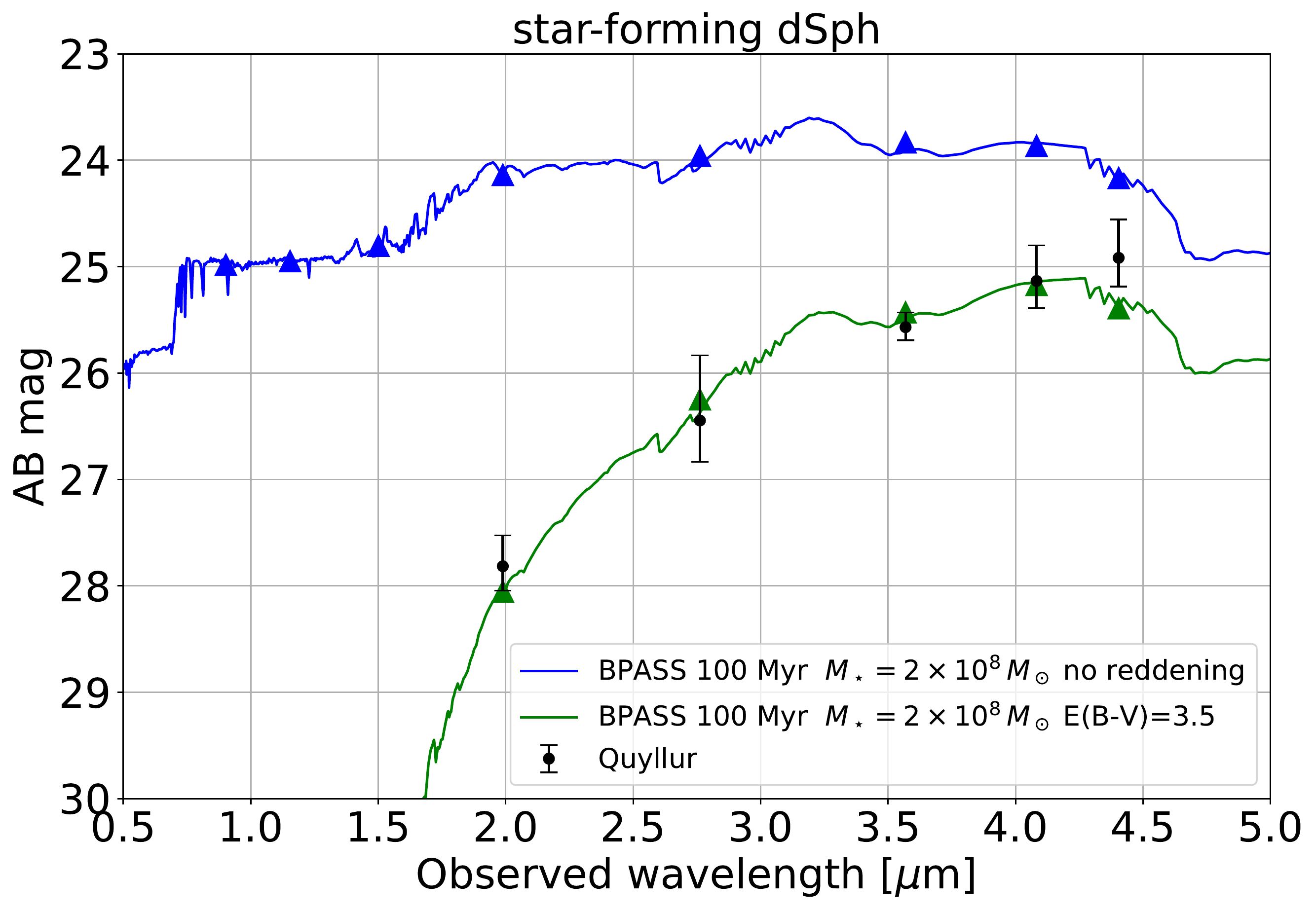}
      \caption{Quyllur compared to SED fits to a single stellar population at El Gordo's redshift $z=0.87$.  Blue lines show the unreddened models as labeled, and green lines show the models reddened with $E(B-V)=3.5$~mag.  Black points with error bars show the Quyllur photometry. For a comparison with the SED of individual stars see Figure~\ref{Fig_Mab_Quillur}. 
         }
         \label{fig:Fig_dSph}
\end{figure}

Figure~\ref{fig:Fig_dSph} shows Quyllur's SED\null. The SED shows a rapid decline below 3~\micron\ consistent with a surface temperature $T\approx 3500$~K at $z=2.1878$, the redshift of 23.1a.
Assuming 30 mas as the maximum separation between counterimages, and accounting for both images (a factor 2 in flux), the magnification must be $\mu>4000$, a boost of at least 9 magnitudes when considering the flux from both images. 
If the separation between images is smaller than 30 mas, then the magnification can be even larger. The observed 3.6~\micron\ magnitude is 25.5, which corresponds to $>$34.5 unmagnified.   At a distance modulus of $\sim$45, and depending on the magnification, the absolute magnitude at 1.1~\micron\ rest wavelength would be ${\sim}-$10.5, consistent with an M2--M4 supergiant.

The required magnification limits the source size and therefore the possible source types.
For $\mu=4000$, the  distance to the caustic must be less than 7.5 microarcseconds or $\approx$0.06~parsec. The only known sources that can satisfy the size and luminosity constraints  are supergiant stars or accretion discs around very massive back holes. 
Some young open clusters with an age around 10--20~Myr show remarkably rich populations of red supergiants \citep{Davies2008RSGC1, Davies2007RSGC2, Alexander2009RSGC3, Froebrich2013}. If the source is not a single red supergiant star, it might be a group of red supergiant stars in an open cluster. This scenario is however disfavored because the constraint on the source size $\lesssim$0.06~pc is significantly smaller than the $\ga$several~pc typical sizes of open clusters. Moreover, red supergiant stars are short-lived, and it is likely that a group of them would be accompanied by  B or even O stars, whose bluer light would be detected unless dust extinction is severe.

The absence of emission shortwards of 1.5~\micron\ (0.47~\micron\ rest) argues against an accretion disk unless there is extinction corresponding to $E(B-V)>4$.  Another argument against an accretion disk is that source-plane reconstruction through the lens model reveals that the source is not located near the nucleus of the host galaxy. 
Photometry performed between Quyllur's position and 23.4a and 23.4b render a color index E(G-J)$\approx 2.5$ (this color index corresponds to the difference between the F410M and F150W filters respectively), while for Quyllur we find a much redder index E(G-J)$\approx 5$. This rules out significant reddening affecting the portion of the galaxy surrounding Quyllur, and being multiply imaged, since only Quyllur shows such extreme color variation. 


The possibility that Quyllur is a transient event showing (at the time of observation) only one of the two lensed images is unlikely based on the expected time delay. Based on the time separation between neighboring knots 23.4a and 23.4b (see time delay column in Table~\ref{tab_arcs}), the time delay between Quyllur and an unseen counterimage must be significantly shorter than 0.28~year because it is closer to the critical curve. (Time delays are inversely proportional to magnification.)
For Quyllur's position, the lens model predicts a magnification  $\mu \ga 30\arcsec/d$, where $d$ is the distance to the critical curve in arcseconds measured along the stretched arc.) In order to observe a transient event, such as a SNe, and not its counterimage, the JWST observation must have taken place  within the first days of the event, which is very unlikely.

Lacking spectroscopic confirmation, we consider the possibility that Quyllur is an interloper galaxy, most likely in the El Gordo cluster. However, Quyllur's SED is remarkably red. A quiescent stellar population even as old as the age of the Universe (6.3~Gyr at $z=0.87$) would appear bluer unless  dust extinction is also present. As shown in Figure~\ref{fig:Fig_dSph}, Quyllur can be consistent with either a 6~Gyr stellar population with stellar mass $M_* \sim 4\times 10^9$~\Msol\  or a 100~Myr stellar population with $M_*\sim 2 \times 10^8$. Either would be of substantial mass but would have to be more compact than $\sim$0.5~kpc, i.e., a dwarf spheroidal (dSph) galaxy. However, either scenario requires dust reddening $E(B-V)=3.5$ (assuming Milky Way like reddening curve with $R_V=3.1$). Without observational evidences for such a galaxy  population inside clusters,  the scenario seems less credible than a caustic event. If Quyllur is an interloper with $z<0.87$ or moderately higher, it would still be unusually red and would require dramatic dust reddening following the same argument. If the redshift is even higher, it would be strongly lensed by El Gordo and show counter-lensed images, which are not found.



\begin{figure} 
      \includegraphics[width=8.5cm]{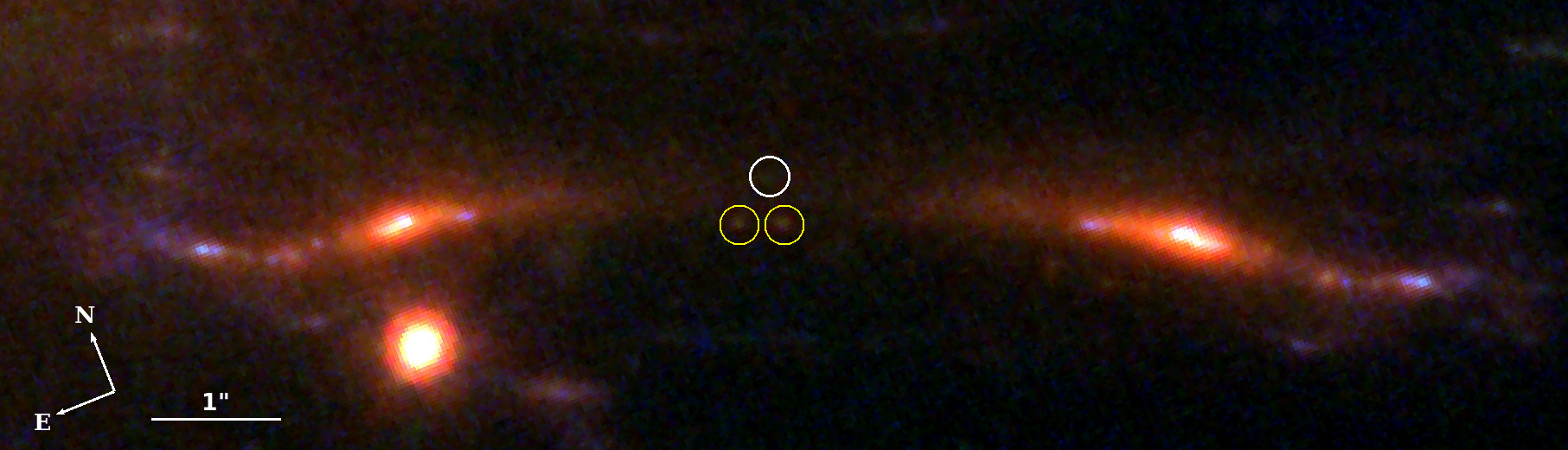}
      \includegraphics[width=8.5cm]{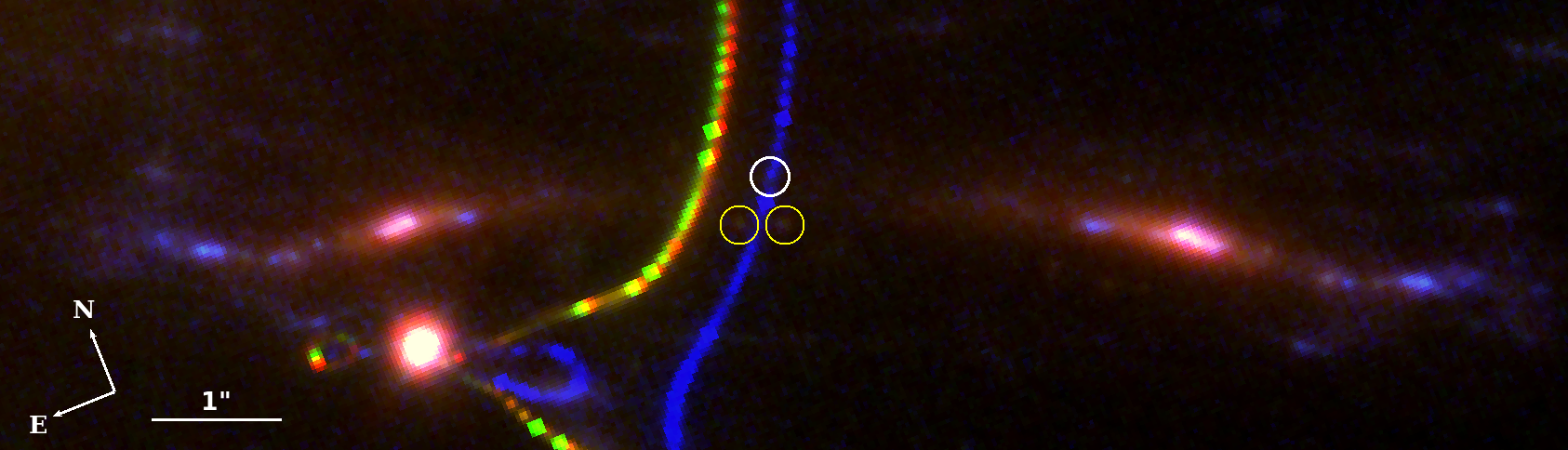}
      \caption{Possible caustic crossing in system~1. {\bf Top:}  color image combining  F090W, F200W and F356W as blue, green, and red, respectively.  {\bf Bottom:} color image in the RGB6 pallette with critical curves from the spectroscopic, Full-r, and Full-a models shown as red, green, and blue, respectively.  In both images,
      the two yellow circles mark images 1.3a and 1.3b, a pair of counterimages close to the expected position of the critical curve. The white circle contains a faint source that is barely resolved. 
         }
         \label{Fig_Star1}
\end{figure}

A second candidate to be a compact source crossing a caustic is within the merging arc of system~1 (Figure~\ref{Fig_Star1}). As before, the two knots 
1.3a and 1.3b are very close to the critical curve and determine the position of the critical curve. A very faint feature is  northwest of this pair and is another candidate for an extremely magnified source no more than a fraction of a parsec from the caustic. The pair of knots 1.3a and 1.3b themselves have  magnification factors $>$100 and are interesting objects to study in more detail.

As discussed in Section~\ref{sect_data}, a possible additional candidate is barely detected in the giant arc La Flaca. However, like the candidate in system~1, it is too faint to extract any useful information. Finally, as shown in Figure~\ref{Fig_AnzueloI}, small, unresolved image pairs can be found in the Anzuelo galaxy. These are not as magnified as Quyllur and hence are likely bigger and brighter sources, such as compact star forming regions or globular clusters. 

\section{Discussion}\label{Sect_Discuss}
In contrast to the lensing cluster WHL J013725.2+140341 \citep{Welch2022a,Welch2022b}, where few new families of lensed galaxies have been identified, El Gordo acts as a magnificent lens amplifying the flux of tens of distant galaxies. This is partially due to the existence of an overdensity at $z\approx 4.3$ \citep[including some of the multiple images already identified in][]{Zitrin2013}, first identified by \cite{Caputi2021} and later confirmed by \cite{Caminha2022} thanks to ALMA and MUSE data respectively, that intersects the caustic region of El Gordo at this redshift.  The dual mass peaks and resulting stretched caustics in El Gordo also help increase its cross section for large magnification factors. In contrast,  earlier lens models of WHL suggest a rounder shape \citep{Welch2022a}, resulting in a  more concentrated caustic region.  These characteristics make El Gordo a good target for lensing studies.

\subsection{Possible tension with $\Lambda$CDM}

The highest mass estimate for El Gordo, 
$M_{\rm vir}\approx 3\times 10^{15}$~\Msol\  \citep[][based on weak lensing]{Jee2014} has significant tension with $\Lambda$CDM\null. Our $M({\rm vir})\approx 2\times10^{15}$ is easier to accommodate, but extrapolating the mass from the region constrained by strong lensing out to the virial radius is less than ideal. Recent work \citep{Asencio2021,Ezquiaga2022} has used weak-lensing limits to illustrate the tension of this cluster with $\Lambda$CDM\null. At the core of this debate is the true mass of El Gordo. 
More recent work (Table~\ref{t:mass}) has tended to give lower masses, but the mass of El Gordo remains uncertain within a factor two. Better weak-lensing data are needed to settle the debate.
Ideally, weak-lensing measurements should be combined with strong-lensing constraints in a self-consistent manner. Strong lensing anchors the mass within the Einstein radius, and weak lensing can extend the mass estimation to beyond the virial radius. The addition of new spectroscopically confirmed strong lensing systems, especially at the highest redshifts, can also help reduce the uncertainty in the mass. Additionally, galaxy--galaxy lensing in the outskirts of El Gordo can be used to constrain the cluster potential, and hence mass, at large radii, because the effective lensing mass of member galaxies scales as their true mass times the large-scale magnification. Constraints on the magnification at a larger radius would constrain the mass within that radius. 

\subsection{Flux ratios} 
\begin{figure} 
      \includegraphics[width=9cm]{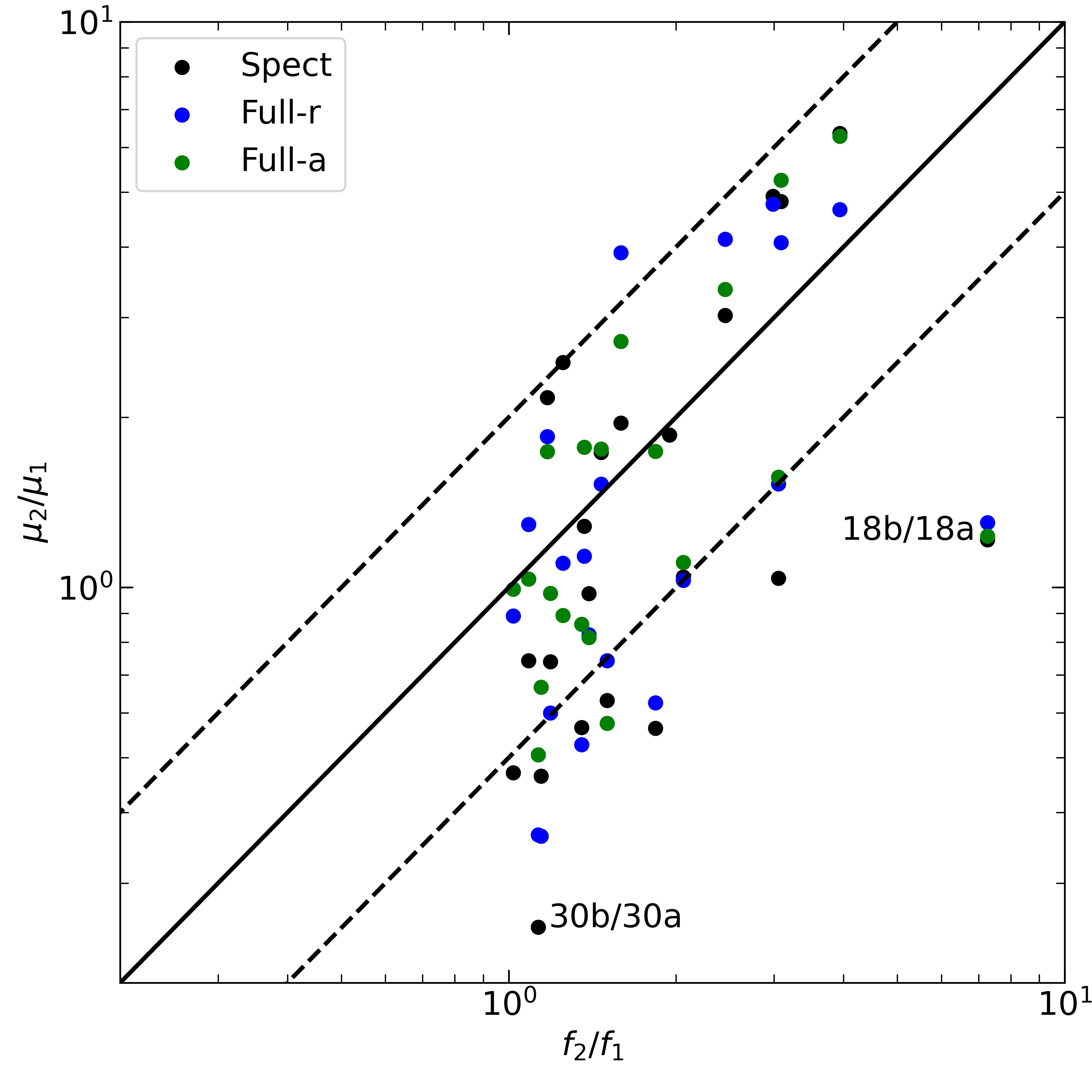}
      \caption{Model magnification ratio versus measured F200W flux ratio for image pairs with more accurate photometry. The flux ratio is defined as the counterimage with the largest flux divided by counterimage with the smallest flux. Solid lines mark equality, and dashed lines mark a factor of two deviation, as expected from uncertainty in the model magnifications. Points are color-coded for each model as shown in the legend, and two outliers are marked. 
         }
         \label{Fig_FluxRatio}
\end{figure}

Comparing the observed flux ratios to the predicted magnification ratios can identify regions where lensing models are inadequate. Differences between the ratios can arise from substructures missing from  the lens model or from transient events affecting some but not all images in a family.  Figure~\ref{Fig_FluxRatio} shows that most images fall within a factor two of expectation, comparable to the expected uncertainty in the lens models. (Uncertainties are typically larger for large magnification factors---\citealt{Zitrin2015,Meneghetti2017}.) 

Only one image family with spectroscopic redshift, System~18, has a flux ratio clearly deviating from the expectations of all three lens models.  The  models predict similar magnification for images 18b and 18a, yet the observed flux of 18b is $\approx$7 times larger than that of 18a. Image 18a has negative parity and is close to a nearby member galaxy, and  its flux may be demagnified by the galaxy. This would require the galaxy to be more massive than predicted by the lens model. (This particular galaxy was not optimized individually but as part of the collection of member galaxies in layer~3 of the lens model.)    
System~30 is also an outlier, in particular for the spectroscopic model that does not use this system as a constraint. The Full-a model optimizes the local mass density, and the model  ratio falls more in line with the observations.

\subsection{Quyllur} Very luminous stars are rarely found isolated and normally live near stellar groups.  The nearby, multiply-lensed pair~23.4a--23.4b is estimated to be at a distance of 10--15~pc from Quyllur, depending on the magnification. This bluer source could be a star-forming region, and Quyllur would be in its outskirts.
Assuming Quyllur is a red supergiant, the magnification is likely significantly higher than our lower limit $\approx$4000. 
Betelgeuse ($d=197$~pc with luminosity  $1.26\times10^5$~\Lsol) 
has a flux density of 
$\approx 3\times 10^{-9}$~ergs~s$^{-1}$~cm$^{-2}$~\AA$^{-1}$ at 6500~\AA\ \citep{Levesque2020}. 
This wavelength would redshift into the F200W JWST filter, for which we find $M_{AB}\approx 28$.  At the distance of Quyllur ($d=1.76\times10^{10}$~pc) and without magnification, Betelgeuse would have an apparent magnitude of $\approx 39.75-2.5*{\rm log}_{10}(1+z)=38.5$ in  F200W (ignoring precise color- or K-correction and extinction). 
This implies the magnification $\approx$15\,000 in order to match Quyllur's apparent magnitude in F200W\null. 
Red hypergiants are rare but have luminosities that can exceed the luminosity of Betelgeuse by a factor of $\approx$ 3--5 (for example UY Scuti and Stephenson 2-18) but still below below the observational limit  \citep{Humphreys1979}. At these luminosities, the required magnification drops to $\approx$4000--7000 (Figure~\ref{Fig_Star1}). 

At magnification factors of several thousand, microlenses are expected to introduce temporary distortions in the flux even for relatively low surface mass density of microlenses ($\Sigma \la 10$~\Msol~pc$^{-2}$). Given the redshift of El Gordo, its intracluster light and  stellar surface mass-density are difficult to measure, but we can approximately estimate the surface mass density of microlenses if we assume the stellar mass contributes $\approx$0.2\%--1\% to the total mass. At the position of Quyllur, the convergence from our lens models  $\kappa=0.63$--0.66, and the critical surface mass-density for the redshifts of El Gordo and Quyllur is 2300~\Msol~pc$^{-2}$. This would make the surface mass-density of microlenses  between 3 and  15~\Msol~pc$^{-2}$. Combined with the model magnification of a few thousand, this implies an effective surface mass density, $\Sigma_{\rm eff}=\mu\times\Sigma$, much larger than the critical surface mass density. In this regime, microlensing events are common \citep{Venumadhav2017, Diego2018}, so we should expect changes in the flux of Quyllur by approximately a factor $\approx 2$ when observing at intervals separated by several months \citep{Welch2022a}.  It is possible that we are observing Quyllur during one of these episodes of increased magnification due to microlenses. Alternatively, red supergiants can exhibit quasi regular changes in their flux, and we could be observing Quyllur during an episode of increased luminosity. 

If Quyllur is confirmed to be a red supergiant, it would be the first example of many to come. So far, all previous stars discovered near a caustic and extremely magnified have  been hot and blue stars. This is partially an observational bias because colder stars, such as red supergiants, at high redshift emit most of their light at wavelengths that are beyond the reach of HST\null.  JWST is most sensitive at these wavelengths \citep{Dai2018}, but red supergiants are intrinsically rarer than blue supergiants because they  correspond to a  shorter-lived stage of massive-star evolution. 

At larger magnification factors, the prospect of detecting the intrinsically fainter red giants at cosmological distances opens the interesting prospect of using them as standard candles (TRGB or Tip of the Red Giant Branch). The tip of the red giant branch is commonly used in our local Universe as part of the distance ladder. 
Extending this ladder up to or beyond $z=1$ would provide an alternative to type Ia supernovae with different (and sometimes easier to manage) systematic effects. The uncertain magnification may be the main obstacle to use these newly found stars in practice although due to the lower luminosity of red giants compared to red supergiants, only those at the most extreme magnification values are expected to be observed. Monitoring for lensing events (that scale as the unknown magnification) will allow to constrain the magnification and open the door to using these stars as standard candles up to $z\approx 1$ and possibly beyond. 

Although not discussed in detail is worth mentioning other exotic objects that show a red SED and could in principle mimic the observed photometry of Quyllur,  including luminous transients that can temporarily reach ${\sim}10^6$~\Lsol\ and can be linked to proto-planetary nebulae \cite{Prieto2009}. 
Finally we can not rule out compact luminous objects such as accretion disks around black holes. We have considered two hypothetical scenarios of obscured AGN that could reproduce the observed colors and magnitudes at redshifts $2.54<z<4.4$. However, at these redshifts we would expect counterimages  with similar colors and fluxes which are not observed.


\begin{figure} 
                \includegraphics[width=8.5cm]{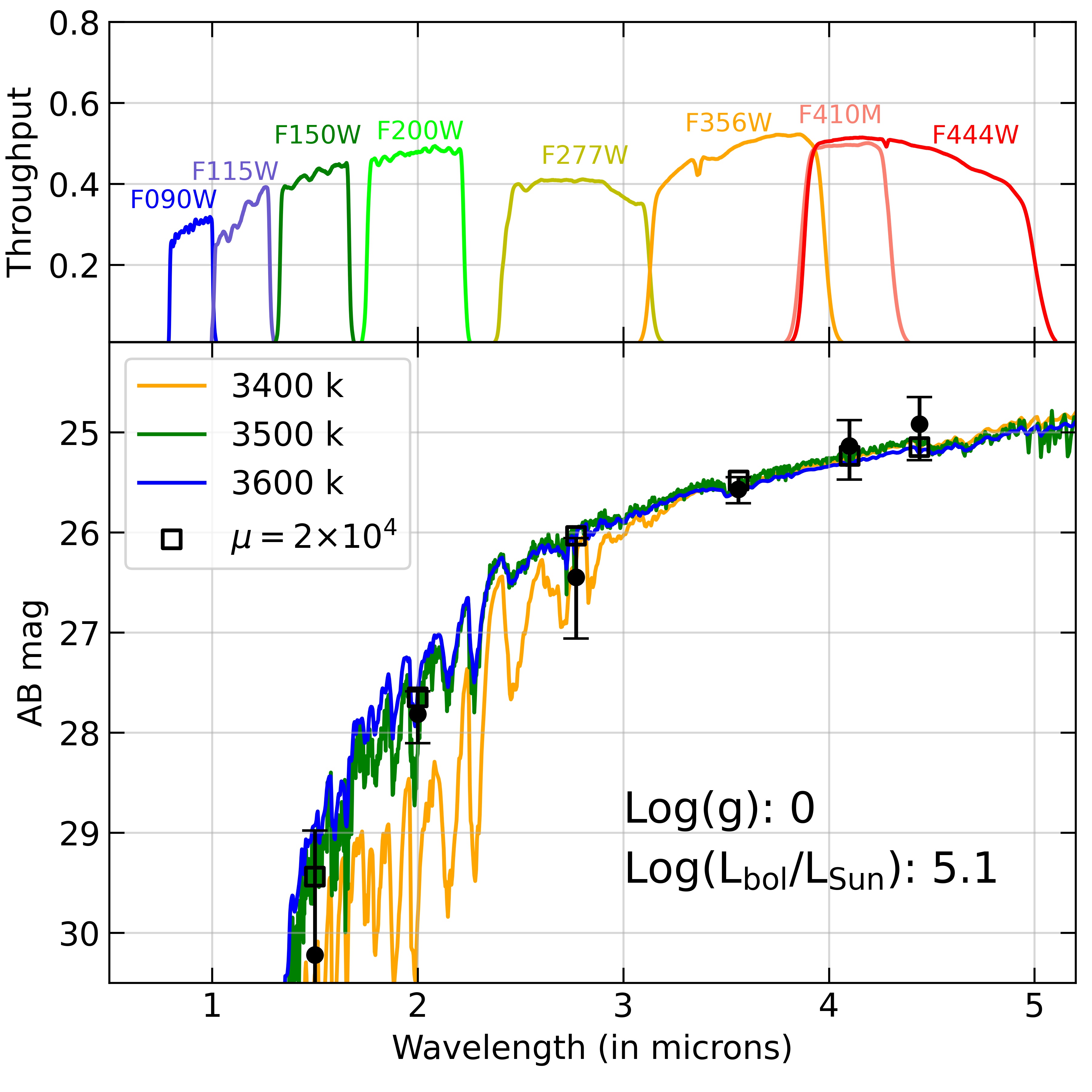}
      \caption{SED of Quyllur compared to stellar models. The black dots show the observed aperture-corrected magnitudes. The blue, green, and orange curves show three models from \cite{Coelho2014} redshifted to $z=2.1878$ and magnified by a factor of 20\,000. Model temperatures are shown in the legend. Squares show the expected flux in the JWST bands for a star with ${ T}\approx 3500$~K and luminosity $10^{5.1}$~\Lsol,  similar to Betelgeuse. A star with similar temperature but a few times brighter, such as UY Scuti, would require magnification less than 10\,000.  
         }
         \label{Fig_Mab_Quillur}
\end{figure}



Regarding the probability of observing a star like Quyllur, one interesting aspect to consider is the large volume accessible through lensing, $V\approx 1.5E11$~Mpc$^3$ between $z=2$ and $z=2.3$ compared with a volume $V\approx 3.3E4$ up to $\approx 20$~Mpc, which is the largest distance up to which we have observed the brightest stars in our local universe. This implies through lensing we can probe a volume that is a factor ${\approx} 5 \times 10^6$ times larger at $2<z<2.3$. A larger volume and at the lower metallicity expected at $z\approx 2$, one should expect to see even brighter stars,  and $\mu \approx 1000$ may suffice to observe them with JWST\null. Only a small fraction of the stars at $z>2$ can be magnified above $\mu\approx 1000$ because $P(\mu>1000) \approx 10^{-8}$ \citep{Diego2019}.
Therefore we would naively expect to see $O(1000)$ stars between $2<z<2.3$ in the entire sky, assuming there are $O(1000)$  stars in our local $r<20$~Mpc volume 
with  $L> 10^5$~\Lsol\
and that these are $\sim$10 times more abundant at $z>2$. 
In the next years, both JWST and HST will continue increasing the number of extremely magnified stars, significantly improving the statistics.

\subsection{Dearth of high-redshift galaxies behind El Gordo}
One of the promises of JWST is the observation of the most distant galaxies. In the case of El Gordo, the existence of overdensities at $z>4$ that fall between the caustics at those redshifts offers  a unique opportunity to study these groups in greater detail \citep{Frye2022}. Our new system candidates  include many  galaxies at high redshift, but the lens models do not place any of them at $z>6$. Because gravitational lenses like El Gordo amplify faint objects, the probability of observing a  $z>6$ galaxy is non-negligible. 
Average magnifications $\mu\ga10$ are possible near the critical curve. Galaxies found in these areas should be stretched in a well-predicted direction. Simply scanning the predicted high-$z$ critical curve, looking for red sources stretched in the expected direction, can easily identify  candidates to be high-redshift galaxies. A search near the $z=10$ critical curve revealed some candidates, but all of them are very faint and are consistent with being stretched arcs at lower redshifts. The  clearest candidate to be a strongly lensed galaxy at high $z$ is at $\rm RA=15.730584,\  Dec=-49.271130$. It is  stretched to a length of $\sim$1\farcs3 in the expected direction, and its position intersects the expected high-redshift critical curves. No other clear candidates were found. For a more detailed search in this cluster using JWST data see \cite{Rachana2022}.

If we assume that there are no highly magnified galaxies at $z>6$ behind El Gordo, we can estimate the likelihood of this lack of galaxies given the lens model and the expected luminosity function at high redshift. The area with total magnification $>$30 is easily computed from the lens model by inverse ray tracing. (Total magnification means after adding all multiple images.) In a typical lensing configuration involving large magnification factors, three images form with two of them carrying most of the magnification, and the third one being significantly less magnified and in some cases falling below the detection limit. (This is the case for some of the systems in table~\ref{tab_arcs}.) We adopt a ratio of 2.5 between the total magnification and the magnification of the brightest counterimage. As shown by \cite{Vega-Ferrero2019}, this is a common ratio found in the Hubble Frontier Fields clusters for systems with large magnification factors. This translates into magnification factors of at least 12 (or 2.7 mag) for the brightest counterimage. 
Our three lens models give areas at $z=6$  with $\mu> 30$ of $\approx$0.015~arcmin$^2$. This is comparable to the area of the six Hubble Frontier Fields clusters, placing El Gordo as the second most efficient lens at high redshift, behind only the train wreck MACS J0717.5+3745 \citep[e.g.,][]{Vega-Ferrero2019}. 

The UV luminosity function \citep{Ishigaki2018} at $z=6$, derived from galaxies behind the Hubble Frontier Fields, reaches  $M_{\rm UV}\approx -16$. A galaxy at $z=6$ (distance modulus 48.85) with absolute magnitude $-$16 and one of its counterimages having magnification 12 would appear at apparent magnitude $\approx 30$, that is, within reach of JWST\null. 
%
%

The volume between redshifts $5.5<z<6.5$ is $V\approx360$~Gpc$^3$ or 2400~Mpc$3$~arcmin${-2}$. 
The volumetric number density of all $5.5<z<6.5$ galaxies with $M_{UV}<-16$ is $N \approx 0.1$ galaxies per Mpc$^{3}$. Hence we expect a surface number density, $N(z=6)=240\, {\rm gal}\, {\rm arcmin}^{-2}$ in the same redshift interval. Most of these galaxies would be out of reach for JWST without  gravitational lensing, but as stated above we expect $\approx$~0.015~arcmin$^2$ behind El Gordo to  be magnified by at least a factor 30, with one counterimage having magnification at least 12. That is, we expect an average of $N \approx 3.6$ galaxies at z=6 strongly magnified, and potentially detectable in the JWST images. 
At higher redshifts the magnification needs to be larger in order to compensate for the increase in distance modulus. The area above a given magnification $\mu$ scales as $\mu^{-2}$, reducing the number of detectable galaxies. Also, the galaxy number density is smaller at higher redshift, so we expect many fewer galaxies at $z\gg6$ than at $z\approx6$.

A deficit of strongly magnified galaxies above $z=6$ could be the result of a patchy reionization scheme, where the UV and visible emission from these galaxies is still being absorbed by a neutral intergalactic medium. Alternative models based on fuzzy dark matter  also predict a lack of galaxies at high redshift due to the existence of a minimum scale, which is determined by the mass of the dark matter particle \citep{Leung2018}. More clusters are needed to improve the statistical significance.

\section{Conclusions}\label{sect_concl}
The El Gordo cluster is a unique object and of great relevance not only to understand galaxy evolution but also for cosmology. With its large mass and redshift, it has been at the center of a debate regarding a possible tension with $\Lambda$CDM\null. 
In the new JWST data we have identified 37 lensed-system candidates (28 of them new ones) which can be added to the 23 spectroscopically confirmed systems from MUSE \citep{Caminha2022} to create three new free-form lens models.
%
The models 
show correlations between the derived mass distribution and the X-ray emission. 
%
Extrapolating from a radius of 500~kpc, where the lens models constrain the mass, to the virial radius of $\sim$2~Mpc gives a mass of about $2.1\times10^{15}$~\Msol, 
close to the limit for standard $\Lambda$CDM cosmology.  Future analyses combining weak and strong lensing will provide observational mass constraints out to the virial radius. 

Interesting individual objects found include  a candidate red supergiant star (nicknamed Quyllur) at $z=2.1878$, and with an estimated magnification of at least 4000. The color of Quyllur is consistent with a red supergiant star, in contrast to other distant, magnified stars, which all appear as blue supergiants. This is in line with the prediction that JWST should greatly outperform HST in the ability to detect highly magnified stars with cool surface temperatures \citep{Dai2018}. Another object of interest is the smallest substructure measured to date at $z= 0.87$ with an estimated mass of $3.8\times10^9$~\Msol, and that creates additional images in portions of  ``La Flaca'', a lensed galaxy 22\arcsec\ long.  Finally, ``El Anzuelo'' is the image of a submillimeter galaxy that can be studied in great detail because of the lensing but it is not modelled in detail here because it still lacks spectroscopic confirmation. 

One anomaly is an apparent deficiency of $z\ga6$ galaxies seen by virtue of El Gordo's magnification.  More systematic searches and quantitative limits are needed.

\section*{acknowledgements}
The authors would like to thank Ignacio Gonzalez-Serrano, and Ignacio Negueruela for useful comments and discussion, and Wilfredo Villca for assisting with the Quechua translation and pronunciation. 
J.M.D. acknowledges the support of project PGC2018-101814-B-100 (MCIU/AEI/MINECO/FEDER, UE) Ministerio de Ciencia, Investigaci\'on y Universidades. This project was funded by the Agencia Estatal de Investigaci\'on, Unidad de Excelencia Mar\'ia de Maeztu, ref. MDM-2017-0765. 
AZ and AKM acknowledge support by Grant No. 2020750 from the United States-Israel Binational Science Foundation (BSF) and Grant No. 2109066 from the United States National Science Foundation (NSF), and by the Ministry of Science \& Technology, Israel.
RAW, SHC, and RAJ acknowledge support from NASA JWST Interdisciplinary
Scientist grants NAG5-12460, NNX14AN10G and 80NSSC18K0200 from GSFC. 
Work by CJC acknowledges support from the European Research Council (ERC) Advanced Investigator Grant EPOCHS (788113). LD acknowledges the research grant support from the Alfred P. Sloan Foundation (Award Number FG-2021-16495). BLF thanks the Berkeley Center for Theoretical Physics for their hospitality during the writing of this paper. MAM acknowledges the support of a National Research Council of Canada Plaskett Fellowship, and the Australian Research Council Centre of Excellence for All Sky Astrophysics in 3 Dimensions (ASTRO 3D), through project number CE17010001.
CNAW acknowledges funding from the JWST/NIRCam contract NASS-0215 to the
University of Arizona.
GBC acknowledge the Max Planck Society for financial support through the Max Planck Research Group for S. H. Suyu and the academic support from the German Centre for Cosmological Lensing.

The authors would like to thank the RELICS team for making the reduced data set available to the community. The scientific results reported in this article are based in part on data obtained from the Chandra Data Archive \footnote{ivo://ADS/Sa.CXO\#obs/12258}\footnote{ivo://ADS/Sa.CXO\#obs/14022}\footnote{ivo://ADS/Sa.CXO\#obs/14023}. 
This work is based on observations made with the NASA/ESA/CSA James Webb Space
Telescope. The data were obtained from the Mikulski Archive for Space
Telescopes at the Space Telescope Science Institute, which is operated by the
Association of Universities for Research in Astronomy, Inc., under NASA
contract NAS 5-03127 for JWST. These observations are associated with JWST
programs 1176 and 2738.

We also acknowledge the indigenous peoples of Arizona, including the Akimel
O'odham (Pima) and Pee Posh (Maricopa) Indian Communities, whose care and
keeping of the land has enabled us to be at ASU's Tempe campus in the Salt
River Valley, where much of our work was conducted.

\software{Astropy: \url{http://www.astropy.org} \citep{Robitaille2013, Astropy2018};
IDL Astronomy Library: \url{https://idlastro.gsfc.nasa.gov} \citep{Landsman1993};
Photutils: \url{https://photutils.readthedocs.io/en/stable/} \citep{Bradley20};
ProFound: \url{https://github.com/asgr/ProFound} \citep{Robotham2017,Robotham2018};
ProFit: \url{https://github.com/ICRAR/ProFit} \citep{Robotham2018};
SExtractor: SourceExtractor:
\url{https://www.astromatic.net/software/sextractor/} or
\url{https://sextractor.readthedocs.io/en/latest/} \citep{Bertin1996}.
}
We would also like to thank Harald Ebeling for making  the code {\small ASMOOTH} \citep{Ebeling2006} available. 

\facilities{ Hubble and James Webb Space Telescope Mikulski Archive
\url{https://archive.stsci.edu}\ }



\newpage

\appendix
\setcounter{table}{0}
\renewcommand{\thetable}{A\arabic{table}}

\section{Compilation of arc positions}\label{Sect_Appendix}
Table~\ref{tab_arcs} lists the complete sample of images used as constraints. 
Systems 1, 2, 5, 10, 23 (images a and b) 28, 29, and 30 were originally used in the lens model of \cite{Zitrin2013}.
Systems 25, 32, and 35 were first added by \cite{Cerny2018} to their lens model. 
Systems 11, 23 (image c), 33, 34, and 36 were first identified and used as additional constraints in \cite{Diego2020}. 
Finally, systems 7 and 8 were identified and used as constraints in \cite{Caputi2021}.  
Systems 1 to 23 are taken directly from \cite{Caminha2022} and use the same ID numbers, which are in column~1. 
Those marked with symbol \dag\ indicate new counterimage candidates found in JWST images in systems that were missing a third counterimage.  RA and Dec positions (FK5) of the counterimages are given in columns~2 and~3 respectively. When available, spectroscopic redshifts from MUSE are shown in column~4. Column~5 lists the geometric redshifts predicted by the lensing model based on spectroscopic redshifts. Geo-$z$ marked with \ddag\ (systems 24 and 53) are unconstrained or poorly constrained by the spectroscopic lens model. System~24 (El Anzuelo) is not included in the set of constraints for the spectroscopic model, and the member galaxies constraining this model are completely unconstrained making impossible any redshift prediction for this system. The assigned redshift corresponds to the photometric redshift derived by \cite{Cheng2022}. The redshift of system~53 is unconstrained by the lens model indicating a possible issue with this system which is close to a spiral member galaxy. This galaxy is modeled together with other members galaxies (ellipticals) and assuming the same light-to-mass ratio which is probably not fully correct. We fix the redshift to z=4 based on the color and location of multiple images of this system. 
Column~6 lists two photometric redshifts derived respectively from the RELICS program (HST) and from the JWST photometry. Images with no photo-$z$ are marked with $-1$. 
Column~7 lists the flux in the F200W band. A $-1$ in this column indicates that there is no detection in this band. Fluxes followed by the symbol \P\ indicate possible systematic errors in the flux estimate due to neighbouring member galaxies or diffraction spikes from bright stars. 
Column~8 lists the magnification predicted by the spectroscopic, Full-r, and Full-a models respectively. The magnification is computed at the corresponding redshift of the system. Magnification values shown as 99.99 are calculated as $>100$, but such high values are unreliable. The magnification is at the exact position given in columns~2 and~3. Changes of a fraction of an arcsecond can result in big changes in the magnification. 
Column~9 lists the time delays predicted by the spectroscopic, Full-r, and Full-a models respectively. The time delay is expressed in years and is relative to the image that arrives first, so by construction is always positive. 
Finally, the column labeled Rank shows the quality of the system. Systems marked with rank A are the most reliable and were used to derive the spectroscopic model. 
Systems marked with B were used to derive (together with systems having rank~A) the Full-r and Full-a models. Counterimages marked with C are less reliable because they cannot be confirmed based on morphology arguments, but they are still  consistent with the spectroscopic lens model.  Counterimages ranked C were not used to constrain any of the models. 
The last three rows with IDs CP1, CP2 and CP3 show the positions used as critical point constrints with their corresponding redshifts. \\

We have checked the MUSE data cube for spectroscopic confirmation of the new system candidates. Only counterimage 55a shows a hint of a line, possibly OII emission at z=1.0714 which would rule out this candidate as strongly lensed. Another faint feature is observed in the same arc and corresponding to H$\delta$ at the same redshift. System 55 is composed of two counterimages separated by only 0.75". Only one of these counterimages shows spectral features in MUSE. 55a is also bluer than 55b. We can not rule out projection effects for being responsible for the spectral features in 55a and bluer color, and as mentioned earlier, this arc could simply not be strongly lensed.  Given the small separation between counterimages 55a and 55b, this arc has very limited constraining power so even if its finally proven not to be strongly lensed, it has a small impact on our lens models. \\

Fluxes in table A1 are obtained from data reduced with Pipeline version 1.6.2 in early August 2022 using the context file jwst\_0942.pmap\_filters. At the time of finalizing this paper a new calibration became available that affects the photometric measurements. Since we relied on geometric redshifts for models Full-r and Full-a, our lens models are insensitive to this change, but fluxes listed in column 8 need to be multiplied by factors, 0.8885, 0.8285, 0.8986, 0.7998 for images falling in modules B1, B2, B3 and B4 respectively in order to reflect the new change in the calibration. The correction factors are smaller at longer wavelengths, with 0.9893,  1.0108,  1.0790, and  1.1128 for the F277W, F356W, F410M, and F444W filters respectively. Since these are the filters in which we detect Quyllur, its corrected  spectrum would be a bit redder ($\approx 0.11$ magnitudes brighter in F444W and $\approx 0.17$ magnitudes fainter in F200W) and would correspond to a slightly cooler star. \\


 \begin{center}  
\begin{longtable}{llLccLLRr@{\hspace{.25em}}r@{\hspace{.25em}}rr@{\hspace{.25em}}r@{\hspace{.25em}}rc}
\caption{Lensed families and images} \label{tab_arcs} \\

\hline \multicolumn{1}{c}{\textbf{ID}} 
& \multicolumn{1}{c}{\textbf{RA}} 
& \multicolumn{1}{c}{\textbf{Dec}} 
& \multicolumn{1}{c}{\textbf{$z_{\rm spec}$}} 
& \multicolumn{1}{c}{\textbf{$z_{\rm geo}$}} 
& \multicolumn{1}{c}{\textbf{$z_{\rm phot}^{{\rm HST}}$}} 
& \multicolumn{1}{c}{\textbf{$z_{\rm phot}^{{\rm JWST}}$}} 
& \multicolumn{1}{c}{\textbf{(MJy/sr)$_{\rm F200W}$}} 
& \multicolumn{3}{c}{\textbf{$\mu$}}
& \multicolumn{3}{c}{\textbf{$\delta T$ (yr)}} 
& \multicolumn{1}{c}{\textbf{Rank}} \\ \hline 
\endfirsthead

\multicolumn{10}{c}%
{{\bfseries \tablename\ \thetable{} -- continued from previous page}} \\
\hline \multicolumn{1}{c}{\textbf{ID}} 
& \multicolumn{1}{c}{\textbf{RA}} 
& \multicolumn{1}{c}{\textbf{DEC}} 
& \multicolumn{1}{c}{\textbf{$z_{\rm spect}$}} 
& \multicolumn{1}{c}{\textbf{$z_{\rm geo}$}} 
& \multicolumn{1}{c}{\textbf{$z_{\rm phot}^{{\rm HST}}$}} 
& \multicolumn{1}{c}{\textbf{$z_{\rm phot}^{{\rm JWST}}$}} 
& \multicolumn{1}{c}{\textbf{(MJy/sr)$_{\rm F200W}$}} 
& \multicolumn{3}{c}{\textbf{$\mu$}} 
& \multicolumn{3}{c}{\textbf{$\delta T$ (yr)}} 
& \multicolumn{1}{c}{\textbf{Rank}} \\ \hline 
\endhead

\hline \multicolumn{8}{r}{{Continued on next page}} \\ \hline
\endfoot

\hline \hline
\endlastfoot
\hline
    1.1a  & 15.722308 & -49.254551 & 2.5635 &   2.64 &  3.46& 2.35 &  201.2 &   7.1&  9.0&  7.8 &  24.56& 22.78& 23.29 & A  \\
    1.1b  & 15.719934 & -49.255207 &  \no   &  \no   &  2.68& 2.33 &  294.4 &  12.3& 13.7& 13.7 &  22.50& 20.94& 21.07 & A  \\
    1.1c  & 15.730896 & -49.250114 &  \no   &  \no   &  3.30& 2.35 &  292.3 &   5.4&  5.6&  6.2 &   0.00&  0.00&  0.00 & A  \\
    1.2a  & 15.722959 & -49.254440 &  \no   &  \no   &  3.31& -1   &    -1  &   8.3& 10.1&  8.8 &  20.22& 18.83& 19.09 & A  \\
    1.2b  & 15.719288 & -49.255486 &  \no   &  \no   &  0.33& -1   &    -1  &   8.9&  9.8&  9.0 &  16.19& 15.33& 15.11 & A  \\
    1.2c  & 15.730638 & -49.250320 &  \no   &  \no   &  3.41& -1   &    -1  &   5.6&  5.9&  6.7 &   0.00&  0.00&  0.00 & A  \\
    1.3a  & 15.721291 & -49.254818 &  \no   &  \no   &   -1 & -1   &    -1  &  99.9& 99.9& 55.9 &   0.00&  0.01&  0.06 & A  \\
    1.3b  & 15.721159 & -49.254852 &  \no   &  \no   &   -1 & -1   &    -1  &  99.9& 99.9& 99.9 &   0.02&  0.00&  0.00 & A  \\
\hline
    2.1a  & 15.733312 & -49.264240 & 2.8254 &   2.77 &  3.25& 2.34 &  208.9 &  25.5& 40.3& 61.4 &   8.16&  6.90&  5.77 & A  \\
    2.1b  & 15.735839 & -49.263081 &  \no   &  \no   &  3.29& -1   &   -1   &  57.3& 91.5& 56.2 &   7.85&  6.48&  5.43 & A  \\
    2.1c  & 15.726700 & -49.267925 &  \no   &  \no   &  3.36& 2.34 &   67.6 &   5.3&  9.9& 11.7 &   0.00&  0.00&  0.00 & A  \\
    2.1d  & 15.736214 & -49.262894 &  \no   &  \no   &  3.40& -1   &  649.9 &  39.7& 61.4& 50.9 &   7.66&  6.21&  5.23 & A  \\
    2.2a  & 15.731842 & -49.264954 &  \no   &  \no   &  3.14& 2.35 &   70.3 &  17.9& 28.7& 38.3 &   6.14&  5.39&  5.21 & A  \\
    2.2b  & 15.738001 & -49.262112 &  \no   &  \no   &  3.15& -1   &   -1   &  12.8& 18.2& 16.8 &   4.10&  3.54&  3.35 & A  \\
    2.2c  & 15.727015 & -49.267693 &  \no   &  \no   &  2.93& -1   &   -1   &   6.1& 11.8& 14.1 &   0.00&  0.00&  0.00 & A  \\
    2.2d  & 15.737608 & -49.262264 &  \no   &  \no   &  3.08& 0.27 &  286.1 &  16.2& 23.1& 27.9 &   4.21&  3.68&  3.29 & A  \\
    2.2e  & 15.737454 & -49.262356 &  \no   &  \no   &  3.08& -1   &   -1   &  18.2& 26.1& 34.5 &   4.31&  3.78&  3.32 & A  \\
    2.3a  & 15.732741 & -49.264557 &  \no   &  \no   &  3.33& 2.81 & 203.7 &  22.2& 36.2& 55.4 &   7.14&  6.03&  5.38 & A  \\
    2.3b  & 15.736686 & -49.262718 &  \no   &  \no   &  -1  & -1   &  -1    &  27.6& 41.2& 38.0 &   6.28&  5.03&  4.25 & A  \\
    2.3c  & 15.726967 & -49.267799 &  \no   &  \no   &  2.93& 2.69 &   69.9 &   5.8& 11.2& 13.4 &   0.00&  0.00&  0.00 & A  \\
    2.4a  & 15.732897 & -49.264477 &  \no   &  \no   &  -1  & -1   &    -1  &  22.6& 36.6& 56.1 &   0.28&  0.20&  0.45 & A  \\
    2.4b  & 15.736458 & -49.262810 &  \no   &  \no   &  3.44& -1   &    -1  &  33.1& 50.3& 43.7 &   0.14&  0.03&  0.00 & A  \\
    2.4c  & 15.736129 & -49.263012 &  \no   &  \no   &  3.40& -1   &    -1  &  47.1& 74.2& 23.6 &   0.00&  0.00&  0.16 & A  \\
    2.5a  & 15.733771 & -49.264046 &  \no   &  \no   &  3.24& -1   &    -1  &  31.7& 49.0& 74.3 &   0.28&  0.44&  0.17 & A  \\
    2.5b  & 15.735603 & -49.263199 &  \no   &  \no   &  3.29& -1   &    -1  &  78.8& 99.9& 99.9 &   0.00&  0.00&  0.00 & A  \\
    2.6a  & 15.731457 & -49.265121 &  \no   &  \no   &  -1  & -1   &    -1  &   8.9& 13.3& 15.8 &   4.76&  4.30&  4.15 & A  \\
    2.6b  & 15.738383 & -49.261936 &  \no   &  \no   &  1.79& -1   &   -1   &  10.5& 15.0& 15.6 &   1.33&  1.08&  1.33 & A  \\
    2.6c  & 15.727132 & -49.267639 &  \no   &  \no   &  -1  & -1   &   -1   &   6.5& 12.5& 15.1 &   0.00&  0.00&  0.00 & A  \\
\hline
    3a  & 15.715284 & -49.248249 & 3.3300 &   3.41 &  -1   & 0.57  &    1.6 &  21.4& 17.4& 10.2 &   3.33&  3.51&  5.63 & A  \\
    3b  & 15.711274 & -49.251835 &  \no   &  \no   &  -1   & -1    &   -1   &   5.6&  5.5&  4.1 &   0.00&  0.00&  0.00 & A  \\
\hline
    4a  & 15.718456 & -49.250854 & 3.3339 &   3.48 &  -1  & -1     &    -1 &   8.6&  9.1&  5.9 &  34.65& 33.76& 36.98 & A  \\
    4b  & 15.715264 & -49.252579 &  \no   &  \no   &  -1  & 0.07   &   3.2 &  11.2& 11.3&  6.3 &  35.29& 34.62& 37.56 & A  \\
 4c\dag & 15.730289 & -49.245262 &  \no   &  \no   &  -1&   -1     &    -1 &   2.6&  2.5&  2.0 &   0.00&  0.00&  0.00 & B  \\
\hline
    5a  & 15.750037 & -49.263752 & 3.5360 &   3.53 &  4.11& 3.25   &   44.0 &   2.6&  2.7&  2.4 &   0.00&  0.00&  0.00 & A  \\
    5b  & 15.730704 & -49.273903 &  \no   &  \no   &  4.06& 3.38   &   71.1 &   9.3&  5.8&  5.7 &  42.62& 38.56& 39.71 & A  \\
    5c  & 15.735929 & -49.268948 &  \no   &  \no   &  0.10& 3.19   &   77.1 &   6.9&  7.5&  5.9 &  39.07& 38.33& 41.15 & A  \\
\hline
    6a  & 15.747679 & -49.265198 & 4.1879 &   4.32 &  -1 & 4.50    &   1.5  &   4.2&  5.1&  4.5 &  56.21& 61.82& 67.66 & A  \\
    6b  & 15.740833 & -49.267590 &  \no   &  \no   &  -1 & 0.61    &   2.3  &   1.1&  3.0&  3.5 &  49.77& 54.62& 59.25 & A  \\
 6c\dag & 15.726980 & -49.275490 &  \no   &  \no   &  -1 & -1      &    -1  &   4.2&  3.3&  2.7 &   0.00&  0.00&  0.00 & A  \\
\hline
    7a  & 15.740075 & -49.254326 & 4.2306 &   4.23 &  -1 & 0.61    &    2.7 &   5.8&  5.5&  5.3 &   0.00&  0.00&  0.00 & A  \\
    7b  & 15.727575 & -49.260315 &  \no   &  \no   &  -1 & 0.63    &    2.2 &   7.2&  6.8&  6.5 &   6.15&  5.07&  6.26 & A  \\
 7c\dag & 15.721050 & -49.263546 &  \no   &  \no   &  -1 & 0.62    &    2.7 &   8.9&  8.2&  7.2 &  13.53& 12.11& 13.70 & A  \\
\hline
    8a  & 15.733409 & -49.251499 & 4.3175 &   4.32 &  4.32& 0.71   &   12.5 &  13.4& 13.7& 14.9 &  28.77& 29.03& 29.48 & A  \\
    8b  & 15.727631 & -49.254570 &  \no   &  \no   &  -1  & -1     &   -1   &  12.8& 12.3& 13.5 &  19.86& 19.49& 20.49 & A  \\
    8c  & 15.713607 & -49.260296 &  \no   &  \no   &  4.39& 0.68   &    9.0 &  10.8& 11.0& 10.3 &   0.00&  0.00&  0.00 & A  \\
\hline
    9.1a  & 15.732421 & -49.252209 & 4.3196 &   4.32 &  4.41& 0.90 &  199.4 &  19.0& 19.3& 19.7 &  39.60& 38.55& 39.47 & A  \\
    9.1b  & 15.728642 & -49.254150 &  \no   &  \no   &  -1  & -1   & 7188\P &   4.8&  5.6&  6.4 &  31.99& 30.61& 31.48 & A  \\
    9.1c  & 15.713000 & -49.260715 &  \no   &  \no   &  4.35& -1   &   -1   &   9.9& 10.1&  9.5 &   0.00&  0.00&  0.00 & A  \\
    9.2a  & 15.731791 & -49.252644 &  \no   &  \no   &  -1  & -1   &  103.6 &  26.9& 26.5& 26.1 &  42.11& 40.50& 40.99 & A  \\
    9.2b  & 15.728818 & -49.254063 &  \no   &  \no   &  -1  & 6.09 &   -1   &   6.0&  6.8&  7.6 &  35.95& 34.21& 34.98 & A  \\
    9.2c  & 15.712775 & -49.260841 &  \no   &  \no   &  4.35& 4.35 &   44.4 &   9.4&  9.6&  9.1 &   0.00&  0.00&  0.00 & A  \\
\hline
    10a & 15.734504 & -49.251942 & 4.3275 &   4.31 &  4.40& 0.69   &  59.2 &  14.3& 14.2& 14.4 &  22.14& 22.78& 22.70 & A  \\
    10b & 15.728204 & -49.255409 &  \no   &  \no   &  4.22&-1      &   3.1\P &  17.7& 14.7& 14.2 &  14.38& 14.08& 14.56 & A  \\
    10c & 15.714784 & -49.260666 &  \no   &  \no   &  4.57& 0.72   &  28.7 &  13.7& 13.8& 13.0 &   0.00&  0.00&  0.00 & A  \\
\hline
    11a & 15.732584 & -49.250099 & 4.3278 &   4.23 &  4.17& 0.69   &  119.4 &  40.6& 26.5& 31.2 &  35.94& 35.95& 37.30 & A  \\
    11b & 15.726316 & -49.253468 &  \no   &  \no   &  -1  & -1     &   -1   &   7.4&  8.0&  8.4 &  24.86& 24.52& 25.62 & A  \\
    11c & 15.712399 & -49.259319 &  \no   &  \no   &  4.33& 4.11   &   20.3 &   7.0&  7.1&  6.5 &   0.00&  0.00&  0.00 & A  \\
\hline
    12a & 15.730983 & -49.247025 & 4.7042 &   4.84 &  4.78& 0.65   &   7.0 &   4.2&  4.0&  3.2 &  33.04& 33.14& 36.86 & A  \\
    12b & 15.722460 & -49.251198 &  \no   &  \no   &  4.60& 0.69   &   4.9 &   3.9&  4.3&  4.0 &  23.37& 23.06& 25.82 & A  \\
    12c & 15.710130 & -49.257271 &  \no   &  \no   &  5.27& -1     &    -1 &  13.8& 20.5& 10.6 &   0.00&  0.00&  0.00 & A  \\
\hline
    13a & 15.726745 & -49.257362 & 4.7528 &   4.64 &  -1  & -1     &    -1 &  10.0&  8.6&  7.8 &   0.00&  0.00&  0.00 & A  \\
    13b & 15.717495 & -49.260963 &  \no   &  \no   &  -1  & 5.62   &   0.6 &  15.8& 15.1& 12.3 &   6.04&  5.07&  5.07 & A  \\
13c\dag & 15.736947 & -49.250027 &  \no   &  \no   &  -1  & -1     &   -1  &  10.1&  8.8&  7.6 &   3.21&  5.08&  4.69 & A  \\
\hline
    14a & 15.739624 & -49.256714 & 4.9486 &   5.21 &  -1  & -1     &   -1   &   5.9&  5.9&  5.1 &  13.87& 14.24& 13.34 & A  \\
    14b & 15.732220 & -49.259888 &  \no   &  \no   &  -1  & -1     &   -1   &   8.2&  6.8&  6.1 &   0.00&  0.00&  0.00 & A  \\
\hline
    15a & 15.729754 & -49.269188 & 4.9770 &   9.92 &  -1  & 0.90   &   2.5 &  61.2& 37.9& 33.7 &  73.19& 78.49& 81.55 & A  \\
    15b & 15.729104 & -49.269657 &  \no   &  \no   &  4.79& 4.92   &   6.4 &  15.4& 99.9& 99.9 &  76.62& 82.00& 84.81 & A  \\
15c\dag & 15.751067 & -49.259571 &  \no   &  \no   &  -1  & -1     &   -1  &   2.5&  3.2&  3.3 &   0.00&  0.00&  0.00 & A  \\
\hline
    16a & 15.733957 & -49.254639 & 5.0880 &   6.56 &  -1  & -1     &   -1   &  24.1& 19.8& 11.9 &   7.32&  7.41&  7.46 & A  \\
    16b & 15.731503 & -49.256031 &  \no   &  \no   &  -1  & 5.71   &   0.6  &  21.9& 27.0& 17.7 &   0.00&  0.00&  0.00 & A  \\
\hline
    17a & 15.710960 & -49.248707 & 5.0929 &   3.97 &  -1  & -1     &    3.7 &   2.1&  2.9&  3.3 &   0.00&  0.00&  0.00 & A  \\
    17b & 15.709747 & -49.249615 &  \no   &  \no   &  4.82&  5.18  &    6.7 &  26.6& 31.9& 23.7 &   1.65&  1.62&  1.74 & A  \\
    17c & 15.709491 & -49.248074 &  \no   &  \no   &  -1  & -1     &    -1  &   1.6&  2.3&  2.8 &   8.82&  8.43&  7.74 & A  \\
\hline
    18a & 15.727793 & -49.259342 & 5.1173 &   4.90 &  1.00& 5.19   &   0.5  &   7.0&  6.3&  5.6 &   0.00&  0.00&  0.00 & A  \\
    18b & 15.719045 & -49.263893 &  \no   &  \no   &  5.17& 5.18   &   3.5  &   8.5&  8.2&  6.9 &  11.26& 10.51& 10.66 & A  \\
18c\dag & 15.740268 & -49.253292 &  \no   &  \no   &  -1  & 5.18   &   0.7  &   7.5&  6.7&  6.6 &   1.24&  2.72&  2.57 & A  \\
\hline
    19a & 15.724179 & -49.261345 & 5.1196 &   5.90 &  5.39& 5.19   &   -1  &  34.0& 46.2& 25.4 &  23.41& 21.18& 21.83 & A  \\
    19b & 15.721812 & -49.262676 &  \no   &  \no   &  5.26& 5.18   &   4.3 &  17.7& 13.9& 12.5 &  30.28& 27.99& 28.46 & A  \\
19c\dag & 15.742441 & -49.253113 &  \no   &  \no   &  -1  & 5.17   &  2.3  &   5.2&  4.3&  3.7 &   0.00&  0.00&  0.00 & A  \\
\hline
    20a & 15.733625 & -49.270912 & 5.4851 &   4.62 &  -1  & 3.84   &   1.3 &  12.8& 11.0& 11.0 &  92.93& 96.52&100.79 & A  \\
    20b & 15.731996 & -49.272797 &  \no   &  \no   &  -1  & 4.92   &   2.4 &  58.0& 23.1& 18.7 & 104.42&108.07&112.01 & A  \\
20c\dag & 15.752936 & -49.262028 &  \no   &  \no   &  -1  & -1     &   -1  &   2.4&  2.3&  2.3 &   0.00&  0.00&  0.00 & B  \\
\hline
    21a & 15.740855 & -49.256954 & 5.5811 &   5.40 &  -1  & -1     &    -1 &   6.9&  6.8&  5.9 &  17.78& 18.31& 17.90 & A  \\
    21b & 15.733271 & -49.260494 &  \no   &  \no   &  5.97& 1.32   &   1.0 &  17.2& 11.2&  9.0 &   0.00&  0.00&  0.00 & A  \\
\hline
    22.1a & 15.750608 & -49.276619 & 5.9521 &   5.95 &  5.74& 4.02 &   2.7 &   4.1&  4.0&  3.8 &   8.24&  8.15&  8.45 & A  \\
    22.1b & 15.755788 & -49.270924 &  \no   &  \no   &  5.73& 4.71 &   3.8 &   4.0&  3.3&  3.1 &   0.00&  0.00&  0.00 & A  \\
    22.2a & 15.751039 & -49.276241 &  \no   &  \no   &  -1  & 4.70 &   1.0 &   5.3&  4.9&  4.7 &   5.13&  4.82&  5.38 & A  \\
    22.2b & 15.755374 & -49.271328 &  \no   &  \no   &  -1  & 6.32 &   1.5 &   5.0&  4.0&  4.0 &   0.00&  0.00&  0.00 & A  \\
\hline
    23.1a & 15.748304 & -49.273941 & 2.1878 &   2.18 &  -1  & -1   &   -1   &  92.3& 83.9& 28.0 &  16.21& 16.29& 18.19 & A  \\
    23.1b & 15.747483 & -49.274567 &  \no   &  \no   &  -1  & -1   &   -1   &  18.8& 18.4& 24.7 &  15.57& 15.66& 17.39 & A  \\
    23.1c & 15.740618 & -49.277542 &  \no   &  \no   &  -1  & -1   & 572.7  &  11.7&  7.4&  5.3 &   0.00&  0.00&  0.00 & A  \\
    23.2a & 15.749012 & -49.273483 &  \no   &  \no   &  -1  & -1   &   -1   &  21.5& 18.3& 12.9 &  13.06& 12.92& 13.59 & A  \\
    23.2b & 15.747150 & -49.274948 &  \no   &  \no   &  -1  & -1   &   -1   &   8.4&  8.9&  9.3 &  12.19& 12.04& 13.19 & A  \\
    23.2c & 15.741710 & -49.277466 &  \no   &  \no   &  -1  & -1   &   -1   &   7.4&  6.5&  5.8 &   0.00&  0.00&  0.00 & A  \\
    23.3a & 15.748508 & -49.274044 &  \no   &  \no   &  -1  & -1   &   -1   &  46.0& 45.5& 23.5 &  12.71& 12.50& 14.01 & A  \\
    23.3b & 15.747520 & -49.274761 &  \no   &  \no   &  1.24& -1   &   -1   &  13.4& 13.9& 16.9 &  11.97& 11.77& 13.23 & A  \\
    23.3c & 15.741501 & -49.277485 &  \no   &  \no   &  3.38& -1   &   -1   &   8.8&  7.1&  6.2 &   0.00&  0.00&  0.00 & A  \\
    23.4a & 15.747970 & -49.274345 &  \no   &  \no   &  2.06& 2.17 & 680.5  &  99.9& 99.1& 99.9 &   0.25&  0.25&  0.28 & A  \\
    23.4b & 15.747764 & -49.274494 &  \no   &  \no   &  2.09& -1   &   -1   &  39.3& 35.4& 75.2 &   0.00&  0.00&  0.00 & A  \\
    23.5a & 15.748455 & -49.273739 &  \no   &  \no   &  1.85& -1   &   -1   &  53.4& 44.8& 20.7 &   0.83&  0.79&  0.98 & A  \\
    23.5b & 15.747318 & -49.274639 &  \no   &  \no   &  -1  &- 1   &   -1   &  14.4& 14.4& 17.4 &   0.00&  0.00&  0.00 & A  \\
\hline
    24.1a & 15.705067 & -49.252438 &  \no & 3.28\ddag &  -1 & -1   &   -1   &   3.0&  3.4& 24.8 &   1.43&  0.67&  1.25 & B  \\
    24.1b & 15.705534 & -49.251270 &  \no   &  \no   &  -1 &  -1   &   -1   &   4.0&  4.2& 15.8 &   1.77&  1.56&  1.60 & B  \\
    24.1c & 15.706085 & -49.251698 &  \no   &  \no   &  -1 &  -1   &   -1   &   4.3&  7.7& 43.9 &   0.00&  0.00&  0.00 & B  \\
    24.2a & 15.704784 & -49.251678 &  \no   &  \no   &  4.18& -1   &   -1   &   3.9& 99.9& 99.9 &   0.31&  0.46&  0.50 & B  \\
    24.2b & 15.704821 & -49.251530 &  \no   &  \no   &  -1  & -1   &   -1   &   4.1& 16.7& 99.9 &   0.67&  0.84&  0.89 & B  \\
    24.2c & 15.705364 & -49.251076 &  \no   &  \no   &  1.00& -1   &   -1   &   4.3&  4.1& 11.3 &   0.00&  0.00&  0.00 & B  \\
    24.3a & 15.704710 & -49.251759 &  \no   &  \no   &  4.18& -1   &   -1   &   3.7& 17.7& 99.9 &   0.16&  0.17&  0.12 & B  \\
    24.3b & 15.704881 & -49.251320 &  \no   &  \no   &  1.19& -1   &   -1   &   4.1&  9.7& 23.4 &   0.05&  0.09&  0.13 & B  \\
    24.3c & 15.705060 & -49.251179 &  \no   &  \no   &  1.19& -1   &   -1   &   4.1&  5.3& 14.0 &   0.00&  0.00&  0.00 & B  \\
    24.4a & 15.705074 & -49.252602 &  \no   &  \no   &  -1  & -1   &   -1   &   2.8&  2.9& 16.0 &   0.09&  0.00&  0.26 & B  \\
    24.4b & 15.705635 & -49.251366 &  \no   &  \no   &  -1  & 4.02 & 1153   &   3.9&  4.7& 20.2 &   0.00&  0.38&  0.00 & B  \\
    24.5a & 15.704923 & -49.252445 &  \no   &  \no   &  1.85& -1   &   -1   &   2.9&  3.3& 20.9 &   1.99&  1.04&  1.88 & B  \\
    24.5b & 15.705960 & -49.251736 &  \no   &  \no   &  -1  & -1   &   -1   &   4.4&  2.6& 22.3 &   0.00&  0.00&  0.00 & B  \\
    24.6a & 15.704995 & -49.252281 &  \no   &  \no   &  -1  & 2.35 &   -1   &   3.1&  5.0& 52.2 &   1.05&  1.18&  1.41 & B  \\
    24.6b & 15.705627 & -49.251217 &  \no   &  \no   &  -1  & -1   &   -1   &   4.1&  3.9& 14.6 &   0.00&  0.00&  0.00 & B  \\
    24.7a & 15.704800 & -49.252190 &  \no   &  \no   &  1.85& -1   & 932.9  &   3.2&  5.5& 61.0 &   2.38&  1.64&  2.26 & B  \\
    24.7b & 15.706110 & -49.251854 &  \no   &  \no   &  -1  & -1   &   -1   &   5.4&  1.3&  9.5 &   0.00&  0.00&  0.00 & B  \\
\hline
    25a & 15.727308 & -49.249607 &  \no   &   2.57 &  3.59& 2.44   &  211.0 &   7.8&  9.6&  9.9 &   4.63&  3.92&  5.62 & B  \\
    25b & 15.721937 & -49.251926 &  \no   &  \no   &  1.06& 2.33   &  156.1 &  13.8& 18.2& 11.5 &  13.55& 12.21& 14.32 & B  \\
    25c & 15.715987 & -49.254715 &  \no   &  \no   &  2.49& 2.43   &  122.3 &   4.2&  4.4&  3.5 &   0.00&  0.00&  0.00 & B  \\
\hline
    26a & 15.751102 & -49.264565 &  \no   &   2.17 &  -1  & -1     &   -1   &   1.9&  1.8&  1.7 &   0.00&  0.00&  0.00 & B  \\
    26b & 15.737077 & -49.270100 &  \no   &  \no   &  -1  & -1     &   -1   &  23.9& 14.3& 10.1 &  65.42& 67.03& 69.11 & B  \\
    26c & 15.732877 & -49.274586 &  \no   &  \no   &  -1  & -1     &   -1   &   4.7&  3.2&  3.2 &  51.95& 49.01& 47.98 & B  \\
\hline
    27a & 15.734736 & -49.271160 &  \no   &   5.00 &  -1  & -1     &   -1   &  11.3&  9.4&  9.2 &   0.00&  0.00&  0.00 & B  \\
    27b & 15.734384 & -49.271507 &  \no   &  \no   &  -1  & -1     &   -1   &   8.3&  7.8&  8.4 &   2.33&  2.39&  2.24 & B  \\
\hline
    28a & 15.731275 & -49.268677 &  \no   &   3.09 &  4.55& 4.14   &   12.1 &  99.9& 46.3& 36.3 &  27.57& 29.95& 32.88 & B  \\
    28b & 15.728963 & -49.270760 &  \no   &  \no   &  4.54& 3.76   &   11.4 &   6.5& 11.6& 12.7 &  27.96& 29.99& 32.01 & B  \\
    28c & 15.747292 & -49.262363 &  \no   &  \no   &  4.96& 0.79   &   11.8 &   3.9&  4.6&  4.6 &   0.00&  0.00&  0.00 & B  \\
\hline
    29a & 15.732704 & -49.268757 &  \no   &   3.28 &  3.26& 3.24   &  29.3  &  20.4& 17.4& 15.2 &  44.06& 45.55& 48.59 & B  \\
    29b & 15.730179 & -49.271206 &  \no   &  \no   &  0.63& 3.16   &  29.8  &   9.6& 15.5& 15.1 &  47.94& 48.90& 50.76 & B  \\
    29c & 15.748834 & -49.262169 &  \no   &  \no   &  -1  & 2.99   &  8.5   &   3.2&  3.6&  3.5 &   0.00&  0.00&  0.00 & B  \\
\hline
    30a & 15.734704 & -49.268829 &  \no   &   2.45 &  2.90& 2.35   &  36.1  &  34.3& 27.1& 16.8 &  24.39& 25.03& 26.67 & B  \\
    30b & 15.731937 & -49.271515 &  \no   &  \no   &  2.28& 2.61   &  40.7  &   8.6&  9.9&  8.5 &  14.62& 14.13& 14.45 & B  \\
    30c & 15.746125 & -49.264801 &  \no   &  \no   &  2.63& 2.55   &  28.5  &   3.9&  4.7&  4.2 &   0.00&  0.00&  0.00 & B  \\
\hline
    31a & 15.741554 & -49.267647 &  \no   &   1.88 &  -1  & -1     &   -1   &  12.1& 75.9& 36.3 &   0.00&  0.00&  0.00 & B  \\
    31b & 15.740235 & -49.267994 &  \no   &  \no   &  -1  & -1     &   -1   &  20.0& 99.9& 37.1 &   3.85&  4.16&  4.71 & B  \\
\hline
    32a & 15.727475 & -49.248367 &  \no   &   2.51 &  3.52& 2.35   &  20.6  &   6.8&  7.5&  8.5 &   0.00&  0.00&  0.00 & B  \\
    32b & 15.721017 & -49.251347 &  \no   &  \no   &  3.24& 2.25   &  15.0  &   5.3&  6.6&  4.8 &  14.02& 13.50& 14.38 & B  \\
    32c & 15.715858 & -49.253979 &  \no   &  \no   &  2.75& 2.35   &  14.2  &   4.4&  4.5&  3.4 &   5.45&  5.60&  3.97 & B  \\
\hline
    33a & 15.749713 & -49.275059 &  \no   &   2.41 &  -1  & -1     &    5.6 &  33.0& 47.9& 38.7 &   0.79&  0.79&  0.32 & B  \\
    33b & 15.748813 & -49.275803 &  \no   &  \no   &  2.92& 1.10   &    2.3 &  10.9& 11.6& 11.5 &   0.00&  0.00&  0.00 & B  \\
\hline
    34a & 15.750683 & -49.262867 &  \no   &   4.06 &  2.84& 1.85   &  13.9  &   2.7&  2.7&  2.5 &   0.00&  0.00&  0.00 & B  \\
    34b & 15.729938 & -49.273075 &  \no   &  \no   &  0.59& 0.32   &   9.4  &  11.3&  8.7&  8.4 &  58.05& 56.91& 59.54 & B  \\
    34c & 15.734825 & -49.269024 &  \no   &  \no   &  -1  & -1     &  36.1  &   7.3&  8.3&  6.7 &  48.09& 48.59& 52.05 & B  \\
\hline
    35a & 15.743917 & -49.276943 &  \no   &   2.22 &  1.21& 0.18   &  12.8  &  38.4& 37.2& 49.1 &   0.00&  0.00&  0.36 & B  \\
    35b & 15.744808 & -49.276600 &  \no   &  \no   &  1.44& 1.78   &   7.0  &  68.1& 59.5& 28.2 &   0.59&  0.51&  0.80 & B  \\
    35c & 15.750566 & -49.272476 &  \no   &  \no   &  3.30& 2.18   &   8.7  &   7.5&  6.3&  6.7 &   1.81&  1.14&  0.00 & B  \\
\hline
    36a & 15.737700 & -49.263165 &  \no   &   2.73 &  -1  & 5.90   &    4.4 &  16.4& 23.0& 28.0 &   8.71&  7.24&  6.26 & B  \\
    36b & 15.732442 & -49.265633 &  \no   &  \no   &  -1  & -1     &   -1   &  27.7& 48.5& 79.3 &   8.71&  7.91&  7.11 & B  \\
    36c & 15.727391 & -49.268688 &  \no   &  \no   &  -1  & 6.65   &    2.4 &   4.7&  9.0& 10.9 &   0.00&  0.00&  0.00 & B  \\
\hline
    37a & 15.732345 & -49.269161 &  \no   &   4.23 &  -1  & 0.64   &    2.0 &  15.2& 14.1& 13.4 &   0.00&  0.00&  0.00 & B  \\
    37b & 15.729484 & -49.272163 &  \no   &  \no   &  -1  & 0.67   &    0.4 &  10.8& 12.5& 12.3 &  11.30& 11.59& 10.80 & B  \\
\hline
    38a & 15.745758 & -49.265537 &  \no   &   3.13 &  0.56& -1     &   -1   &   5.5&  8.0&  7.9 &  56.77& 57.80& 60.01 & B  \\
    38b & 15.744725 & -49.265804 &  \no   &  \no   &  3.92& -1     &   -1   &  17.8& 37.6& 55.8 &  57.02& 58.34& 59.91 & B  \\
    38c & 15.744463 & -49.265915 &  \no   &  \no   &  3.73& 2.85   &  60.8  &  16.6& 36.5& 53.0 &  56.54& 57.95& 59.60 & B  \\
    38d & 15.740879 & -49.266903 &  \no   &  \no   &  -1  & -1     &   -1   &   3.5& 28.1& 47.2 &  56.09& 57.14& 58.62 & B  \\
    38e & 15.728203 & -49.274441 &  \no   &  \no   &  3.30& -1     &    -1  &   5.0&  3.9&  3.4 &   0.00&  0.00&  0.00 & B  \\
\hline
    39a & 15.729329 & -49.250839 &  \no   &   2.47 &  -1  & 0.71   &    8.1 &   6.9&  7.6&  8.7 &   0.00&  0.00&  0.00 & B  \\
    39b & 15.726809 & -49.252319 &  \no   &  \no   &  4.49& 4.11   &    3.6 &  32.3& 37.5& 23.3 &   6.43&  6.41&  6.84 & B  \\
    39c & 15.718955 & -49.255386 &  \no   &  \no   &  0.53& -1     &    -1  &   7.5&  8.1&  7.3 &   3.76&  4.27&  4.98 & B  \\
\hline
    40a & 15.735991 & -49.263199 &  \no   &   2.75 &  -1  & -1     &   -1   &  41.0& 56.5& 44.4 &   7.29&  6.19&  5.46 & B  \\
    40b & 15.733200 & -49.264481 &  \no   &  \no   &  -1  & -1     &  208.9 &  29.5& 53.8& 99.9 &   7.62&  6.48&  5.70 & B  \\
    40c & 15.727150 & -49.267715 &  \no   &  \no   &  -1  & -1     &   69.9 &   6.0& 11.3& 13.5 &   0.00&  0.00&  0.00 & B  \\
\hline
    41a & 15.714717 & -49.251659 &  \no   &   4.64 &  5.68& 4.71   &   3.4  &  99.9& 99.9& 25.6 &   0.98&  0.93&  0.88 & B  \\
    41b & 15.715283 & -49.251335 &  \no   &  \no   &  5.64& 6.14   &   4.2  &  43.1& 32.4& 80.2 &   0.00&  0.00&  0.00 & B  \\
\hline
    42a & 15.741546 & -49.266411 &  \no   &   2.26 &  2.32& 2.28   &   10.7 &  12.3& 24.6& 15.8 &  30.62& 27.86& 29.72 & B  \\
    42b & 15.740117 & -49.266598 &  \no   &  \no   &  -1  & -1     &  198.8\P &   3.2& 12.2& 17.9 &  34.37& 31.38& 32.62 & B  \\
    42c & 15.731348 & -49.272121 &  \no   &  \no   &  -1  & 2.34   &    6.7 &   6.3&  6.3&  5.8 &   0.00&  0.00&  0.00 & B  \\
\hline
    43a & 15.728325 & -49.266068 &  \no   &   3.75 &  -1  & 3.33   &  0.9  &  13.7& 14.6& 14.5 &  23.64& 24.81& 24.27 & B  \\
    43b & 15.725151 & -49.267990 &  \no   &  \no   &  -1  & -1     &  -1   &   4.8&  9.4& 11.2 &  28.40& 29.54& 29.48 & B  \\
    43c & 15.745181 & -49.258690 &  \no   &  \no   &  -1  & -1     &  -1   &   4.3&  4.7&  5.1 &   0.00&  0.00&  0.00 & C  \\
\hline
    44a & 15.730420 & -49.262566 &  \no   &   3.11 &  -1  & -1     &  65.8 &  11.7& 12.9& 14.4 &   2.75&  2.34&  1.76 & B  \\
    44b & 15.732507 & -49.261646 &  \no   &  \no   &  -1  & -1     &   4.2 &  37.7& 27.2& 20.7 &   2.22&  1.85&  1.44 & B  \\
    44c & 15.724905 & -49.264732 &  \no   &  \no   &  -1  & -1     &   -1  &  12.2& 12.0& 12.4 &   0.00&  0.00&  0.00 & C  \\
\hline
    45a & 15.734213 & -49.258305 &  \no   &   4.88 &  -1  & 0.21   &   1.5 &  14.8& 15.9& 12.7 &   0.00&  0.00&  0.00 & B  \\
    45b & 15.734550 & -49.258129 &  \no   &  \no   &  -1  & -1     &   -1  &  22.5& 25.6& 19.9 &   0.98&  1.06&  1.03 & B  \\
    45c & 15.734900 & -49.257996 &  \no   &  \no   &  0.66& -1     &   -1  &  99.9& 99.9& 99.9 &   1.71&  1.83&  1.86 & B  \\
\hline
    46a & 15.744821 & -49.279274 &  \no   &   5.86 &  -1  & 7.21   &   1.0 &  20.1& 36.1& 14.7 &   0.00&  0.00&  0.00 & B  \\
    46b & 15.743759 & -49.279541 &  \no   &  \no   &  -1  & 2.35   &   2.1 &  56.1& 23.1&  6.4 &   1.19&  1.13&  0.80 & B  \\
\hline
    47a & 15.713725 & -49.250370 &  \no   &   2.78 &  -1  & -1     &   -1   &  37.8& 84.5& 99.9 &   0.07&  0.06&  0.40 & B  \\
    47b & 15.712871 & -49.250942 &  \no   &  \no   &  -1  & 4.02   &   2.4  &   9.2&  9.1&  6.7 &   0.00&  0.00&  0.00 & B  \\
\hline
    48a & 15.727625 & -49.246033 &  \no   &   3.15 &  -1  & 2.79   &   6.3  &   4.8&  4.6&  4.2 &   7.36&  7.80& 13.43 & B  \\
    48b & 15.712209 & -49.253826 &  \no   &  \no   &  -1  & 0.17   &   5.1  &   3.7&  3.6&  3.0 &   0.00&  0.00&  0.00 & B  \\
\hline
    49a & 15.721458 & -49.254242 &  \no   &   2.44 &  -1  & -1     &   2.6  &  13.8& 19.0& 12.6 &  25.84& 23.83& 24.27 & B  \\
    49b & 15.719934 & -49.254555 &  \no   &  \no   &  -1  & 3.80   &   3.1  &  10.2& 11.4& 12.3 &  23.79& 21.91& 22.46 & B  \\
    49c & 15.729680 & -49.250084 &  \no   &  \no   &  -1  & -1     &   3.0  &   6.4&  6.9&  8.4 &   0.00&  0.00&  0.00 & B  \\
\hline
    50a & 15.732396 & -49.257103 &  \no   &   2.46 &  1.00& -1     &   -1   &  14.2& 12.1& 12.5 &   0.00&  0.00&  0.00 & B  \\
    50b & 15.731825 & -49.257393 &  \no   &  \no   &  -1  & -1     &   -1   &  21.8& 18.6& 22.3 &   0.99&  0.98&  0.94 & B  \\
    50c & 15.728638 & -49.258949 &  \no   &  \no   &  -1  & -1     &   -1   &  20.0& 19.1& 22.3 &   3.36&  3.48&  3.59 & B  \\
    50d & 15.725208 & -49.260342 &  \no   &  \no   &  -1  & -1     &   -1   &  14.4& 13.2& 12.5 &   1.22&  1.68&  1.63 & B  \\
\hline
    51a & 15.722816 & -49.253960 &  \no   &   2.61 &  0.38& 2.06   &   4.7   &   3.6&  4.6&  4.2 &  21.22& 19.55& 20.14 & B  \\
    51b & 15.718608 & -49.255116 &  \no   &  \no   &  -1  & 2.17   &   5.6   &   7.8&  8.5&  7.3 &  16.31& 15.17& 14.62 & B  \\
    51c & 15.730049 & -49.249874 &  \no   &  \no   &  1.80& 2.54   &   3.2   &   5.7&  6.2&  7.3 &   0.00&  0.00&  0.00 & B  \\
\hline
    52a & 15.732757 & -49.264423 &  \no   &   2.77 &  3.33& -1     &  -1     &  26.0& 45.5& 80.9 &   8.10&  6.86&  6.37 & B  \\
    52b & 15.736487 & -49.262676 &  \no   &  \no   &  3.44& -1     &  -1     &  24.2& 34.0& 35.0 &   7.85&  6.42&  5.58 & B  \\
    52c & 15.726746 & -49.267796 &  \no   &  \no   &  2.93& -1     &  -1     &   5.3&  9.8& 11.5 &   0.00&  0.00&  0.00 & B  \\
\hline
    53a & 15.743875 & -49.260933 &  \no   & 4\ddag &  -1  & -1     &   -1    &   9.7& 12.6& 17.4 &   0.00&  0.00&  0.00 & B  \\
    53b & 15.746392 & -49.261173 &  \no   &  \no   &  -1  & -1     &   -1    &   7.4& 10.0& 14.4 &   5.70&  5.64&  5.64 & B  \\
    53c & 15.744904 & -49.260807 &  \no   &  \no   &  -1  & -1     &   -1    &  11.5& 12.3& 15.3 &   2.52&  2.61&  2.42 & C  \\
\hline
    54a & 15.726033 & -49.268112 &  \no   &   3.80 &  -1  & 0.70   &   5.2   &   5.9& 13.6& 17.8 &  34.49& 35.34& 34.81 & B  \\
    54b & 15.729225 & -49.266129 &  \no   &  \no   &  4.24& 0.66   &   4.6   &   9.9& 11.6& 11.6 &  29.02& 29.68& 29.17 & B  \\
    54c & 15.746192 & -49.258694 &  \no   &  \no   &  0.44& 1.11   &   3.8   &   3.7&  4.2&  4.6 &   0.00&  0.00&  0.00 & B  \\
\hline
    55a & 15.719779 & -49.255875 &  \no   &   4.34 &  0.67& 1.31   &   5.3   &  99.9& 99.9& 99.9 &   0.23&  0.23&  0.07 & B  \\
    55b & 15.720067 & -49.255791 &  \no   &  \no   &  0.91& 0.80   &  14.7   &  99.9& 99.9& 73.8 &   0.00&  0.00&  0.00 & B  \\
\hline
    56a & 15.746338 & -49.265072 &  \no   &   3.12 &  3.17& 3.27   &  248.7  &   5.5&  6.4&  5.8 &  39.75& 42.02& 45.49 & B  \\
    56b & 15.739883 & -49.266621 &  \no   &  \no   &  2.63& 3.51   &  198.8  &   2.2&  5.8&  6.5 &  40.59& 43.34& 46.14 & B  \\
    56c & 15.728309 & -49.274208 &  \no   &  \no   &  3.06& 2.41   &   81.4  &   5.3&  4.2&  3.7 &   0.00&  0.00&  0.00 & B  \\
    56d & 15.740427 & -49.267254 &  \no   &  \no   &  0.63& -1     & 7530\P  &   0.1&  0.5&  0.6 &  42.72& 44.52& 47.47 & B  \\
\hline
    57a & 15.748486 & -49.265724 &  \no   &   3.56 &  -1  & 6.08   &  0.3   &   3.4&  3.8&  3.6 &   4.07&  4.18&  3.81 & B  \\
    57b & 15.743563 & -49.267609 &  \no   &  \no   &  -1  & 8.54   &  1.2   &  21.6& 17.7& 22.6 &   0.00&  0.00&  0.00 & B  \\
\hline
    58a & 15.746034 & -49.278336 &  \no   &   2.85 &  -1  & -1     &   -1   &  60.5& 46.1& 13.1 &   0.00&  0.00&  0.00 & B  \\
    58b & 15.745462 & -49.278591 &  \no   &  \no   &  -1  & -1     &   -1   &  20.5& 17.8&  7.8 &   0.25&  0.28&  0.35 & B  \\
\hline
    59a & 15.719483 & -49.255890 &  \no   &   3.86 &  -1  & 4.72   &  0.8   &  30.7& 41.9& 38.6 &  39.46& 37.12& 39.35 & B  \\
    59b & 15.719833 & -49.255745 &  \no   &  \no   &  0.91& 0.60   &  1.6   &  57.1& 99.9& 99.9 &  39.05& 36.67& 39.07 & B  \\
    59c & 15.735154 & -49.247620 &  \no   &  \no   &  -1  & 0.04   &  2.2   &   3.2&  2.9&  2.3 &   0.00&  0.00&  0.00 & C  \\
\hline  
    60a & 15.724298 & -49.254860 &  \no   &   2.63 &  -1  & 2.32   &  3.3   &  11.9& 13.6& 13.4 &   4.51&  4.10&  4.93 & B  \\
    60b & 15.719618 & -49.256420 &  \no   &  \no   &  -1  & 0.04   &  5.0   &   9.8& 10.8& 11.0 &   0.00&  0.00&  0.00 & B  \\
\hline
\end{longtable}
\tablecomments{ID's for the first 23 systems follow the naming scheme of \cite{Caminha2022}. Counterimages marked with $^{\dag}$ are  new candidates to be third counterimages identified with JWST.}
\end{center}

\newpage 


\bibliographystyle{aasjournal} 
\bibliography{MyBiblio} 


\end{document}